\documentstyle{mn}
\input epsf

\def\pageoffset#1#2{\hoffset=#1\relax\voffset=#2\relax}

\newcommand\qvec{{\bf q}}
\newcommand\xvec{{\bf x}}
\newcommand\vvec{{\bf v}}
\newcommand\rvec{{\bf r}}
\newcommand\uvec{{\bf u}}

\def\spose#1{\hbox to 0pt{#1\hss}}
\def\simlt{\mathrel{\spose{\lower 3pt\hbox{$\mathchar"218$}}
     \raise 2.0pt\hbox{$\mathchar"13C$}}}
\def\simgt{\mathrel{\spose{\lower 3pt\hbox{$\mathchar"218$}}
     \raise 2.0pt\hbox{$\mathchar"13E$}}}
\def\beq{\begin{equation}}
\def\eeq{\end{equation}}
\def\bce{\begin{center}}
\def\ece{\end{center}}
\def\bea{\begin{eqnarray}}
\def\eea{\end{eqnarray}}
\def\ben{\begin{enumerate}}
\def\een{\end{enumerate}}

\def\nn{\nonumber}

\def\brr{\begin{array}}
\def\err{\end{array}}

\def\nh1{n_{\rm HI}}

\def\p1dk{P_{\rm 1D}(k)}
\def\simlt{\mathrel{\spose{\lower 3pt\hbox{$\mathchar"218$}}
     \raise 2.0pt\hbox{$\mathchar"13C$}}}

\def\Or{{\cal O}}

\newcommand{\lexp}{\mathop{\bigl\langle}}

\newcommand{\rexpc}{\mathop{\bigr\rangle_c}{}}

\pageoffset{0pc}{-2.5pc}

\newbox\grsign \setbox\grsign=\hbox{$>$} \newdimen\grdimen \grdimen=\ht\grsign
\newbox\simlessbox \newbox\simgreatbox
\setbox\simgreatbox=\hbox{\raise.5ex\hbox{$>$}\llap
     {\lower.5ex\hbox{$\sim$}}}\ht1=\grdimen\dp1=0pt
\setbox\simlessbox=\hbox{\raise.5ex\hbox{$<$}\llap
     {\lower.5ex\hbox{$\sim$}}}\ht2=\grdimen\dp2=0pt

\title{The Real and Redshift Space Density Distribution Function
for Large-Scale Structure in the Spherical Collapse Approximation}

\author[Scherrer and Gazta\~naga]{Robert J. Scherrer$^1$ and
Enrique Gazta\~naga$^{2,3}$\\
$^1$Department of Physics and Department of Astronomy,
Ohio State University, Columbus, OH 43210\\
$^2$ INAOE, Astrofisica, Tonantzintla, Apdo Postal 216 y 51, 
 Puebla 7200, Mexico \\
$^3$Institut d'Estudia Espacils de Catalunya, Research Unit (CSIC),
Edf. Nexus-104-c/Gran Capita 2-4, 08034 Barcelona, Spain}

\begin{document}

\maketitle
 
\begin{abstract}
  We use the spherical collapse (SC) approximation to derive expressions for
  the smoothed redshift-space probability distribution function (PDF), as well
  as the $p$-order hierarchical amplitudes $S_p$, in both real and redshift
  space.  We compare our results with numerical simulations, focusing on the
  $\Omega=1$ standard CDM model, where redshift distortions are strongest.  We
  find good agreement between the SC predictions and the numerical PDF in real
  space even for $\sigma_L \simgt 1$, where $\sigma_L$ is the linearly-evolved
  rms fluctuation on the smoothing scale.  In redshift space, reasonable
  agreement is possible only for $\sigma_L \simlt 0.4$.  Numerical simulations
  also yield a simple empirical relation between the real-space PDF and
  redshift-space PDF: we find that for $\sigma < 1$, the redshift space
  PDF, $P\left[\delta_{(z)}\right]$, is, to a good approximation, a simple
  rescaling of the real space PDF, $P\left[\delta\right]$, i.e.,
  $P\left[\delta/\sigma\right]d\left[\delta/\sigma\right] =
  P\left[\delta_{(z)}/\sigma_{(z)}\right]
  d\left[\delta_{(z)}/\sigma_{(z)}\right]$, where $\sigma$ and $\sigma_{(z)}$
  are the real-space and redshift-space rms fluctuations, respectively.  This
  result applies well beyond the validity of linear perturbation theory, and
  it is a good fit for both the standard CDM model and the $\Lambda$CDM model.
  It breaks down for SCDM at $\sigma \approx 1$, but provides a good fit
  to the $\Lambda$CDM models for $\sigma$ as large as 0.8.

\end{abstract}


\section{Introduction}
One of the central statistical quantities of interest in the
study of large-scale structure is $P[\rho]$, the one-point probability
distribution function (PDF) of the density field $\rho$.  In the linear
regime, the PDF simply retains its Gaussian shape, with the
variance scaling up as the square of the growth factor.
In the nonlinear regime, however, $P[\rho]$ diverges
away from its (presumably Gaussian) initial shape.

A number of recent studies have been undertaken to predict analytically the
evolution of the PDF.  Kofman et al. (1994) calculated the PDF produced by the
Zeldovich approximation.  Given a formalism to calculate the cumulants of the
evolved distribution (Peebles 1980; Fry 1984; Bernardeau 1992a, hereafter
B92), the Edgeworth expansion can be employed to derive an approximate
expression for the PDF (Juszkiewicz et al. 1995, Bernardeau \& Kofman 1995).
A better alternative is to derive a similar expansion around the Gamma PDF
(Gazta\~naga, Fosalba \& Elizalde 2000).  Unlike the Edgeworth
expansion, the Gamma expansion always yields positive densities.
The generating function for these
cumulants can be used to derive the Fourier transform of the PDF, which can be
inverted to give the PDF in the form of an integral (Bernardeau 1994).
Alternatively, this generating function can be transformed into a local
Lagrangian mapping which, when applied to the initial Gaussian density field,
gives a PDF with the desired tree-level hierarchical amplitudes (Protogeros \&
Scherrer 1997, hereafter PS; Protogeros, Melott, \& Scherrer 1997).  Fosalba
and Gazta\~naga (1998a, hereafter FG) showed that when shear is neglected, the
equations of motion automatically produce a local Lagrangian mapping which
exactly reproduces the tree-level hierarchical amplitudes (also Fosalba and
Gazta\~naga 1998b, Gazta\~naga \& Fosalba 1998).  It is this approach,
neglecting shear in the cosmological equations of motion, which we will
utilize here.  Fosalba \& Gazta\~naga dubbed their approximation the
``spherical collapse" (SC) approximation; although we adopt their terminology,
it would perhaps be more accurate to refer to this approximation as
``shear-free collapse". The SC model has also been used to model the
PDF of QSO Ly-$\alpha$ absorbers (Gazta\~naga \& Croft 1999).

Of course, the three-dimensional distribution of galaxies is observed in redshift
space rather than real space, with the radial coordinate distorted
by the peculiar velocities of the galaxies.  Hence, the most physically
relevant quantity is not the PDF in real space, but the redshift-space
PDF.  While the evolution of the hierarchical amplitudes
in redshift space has been examined in some detail
(Bouchet et al. 1995; Hivon et al. 1995; Scoccimarro et al. 1999),
comparatively
less work has been done on the distortion of the full PDF
in redshift space. 
Kaiser (1987) showed that, to lowest order, the rms fluctuation
in redshift space, $\sigma_{(z)}$, is related to the real-space rms
fluctuation $\sigma$, via
\begin{equation}
\label{kaiser}
\sigma_{(z)}^2 = \left[ 1 + {2 \over 3} f_{\Omega}+ {1 \over 5} f_{\Omega}^2 \right]~\sigma^2
\label{Eq:Kaiser}
\end{equation}
where $f_\Omega \sim \Omega^{0.6}$.
(Eq. 1 assumes that the bias parameter is unity.  We make this
assumption throughout, although bias can be incorporated
into our formalism if the form of the bias is prescribed).
The effect of redshift distortions on the skewness was calculated
by Bouchet et al. (1995) and Hivon et al. (1995).  They found that
redshift distortions have little effect on the skewness.
For $\Omega = 1$, for instance, the redshift-space
skewness is well-approximated by (Hivon et al. 1995)
$S_{3(z)} = 35.2/7 - 1.15(n+3)$; this compares
with a real-space result of
$S_3 = 34/7 - (n+3)$.  (This theoretical result for the redshift-space
skewness does not agree with numerical simulations except at
very large length scales; see, e.g., Fig. 3 below).
A systematic method to calculate the
higher-order perturbation theory kernels in redshift space
has been given by Scoccimarro, Couchman, \& Frieman (1999).
The most detailed treatments thus far
have been given by Hui, Kofman, and Shandarin (2000), who
calculated the redshift-space PDF in the context of the Zeldovich
approximation, and by Watts and Taylor (2000) who calculated
the redshift-space PDF in second-order perturbation theory.

In the next section, we derive the local Lagrangian mapping corresponding to
the SC model, and apply it to Gaussian initial conditions to derive the
real-space density distribution function. In Section 3, we derive an
approximation to the redshift distortion of the density field based on the SC
approximation.  We calculate the corresponding redshift-space hierarchical
moments and the redshift-space PDF in the SC approximation and compare the PDF
to numerical simulations in redshift space.  We also derive an interesting
empirical relation between the real-space PDF and the redshift-space PDF.  Our
conclusions are summarized in Section 4.

\begin{figure*}
\centering\centerline{
\epsfxsize=5.7cm \epsfbox{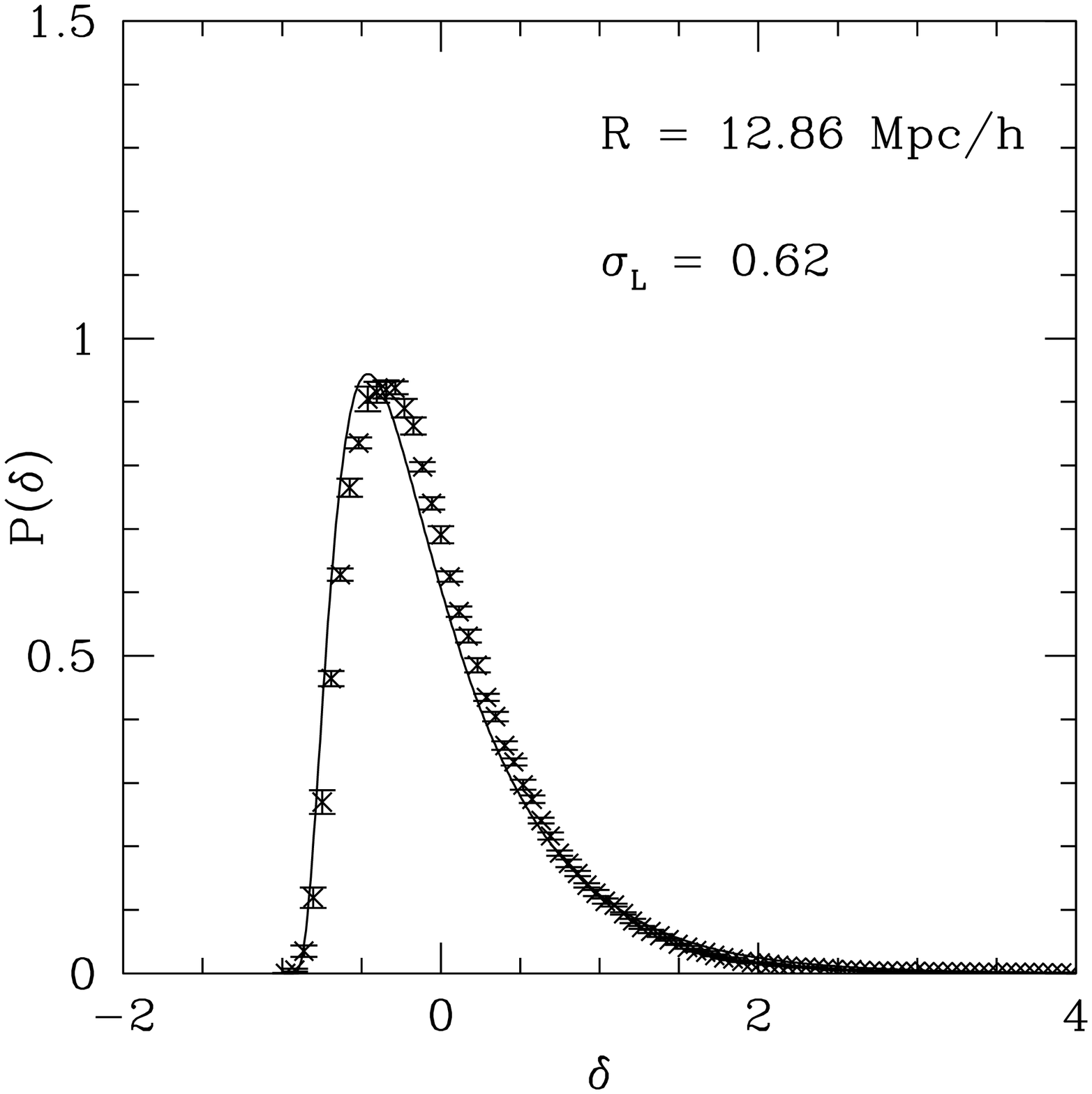}
\epsfxsize=5.7cm \epsfbox{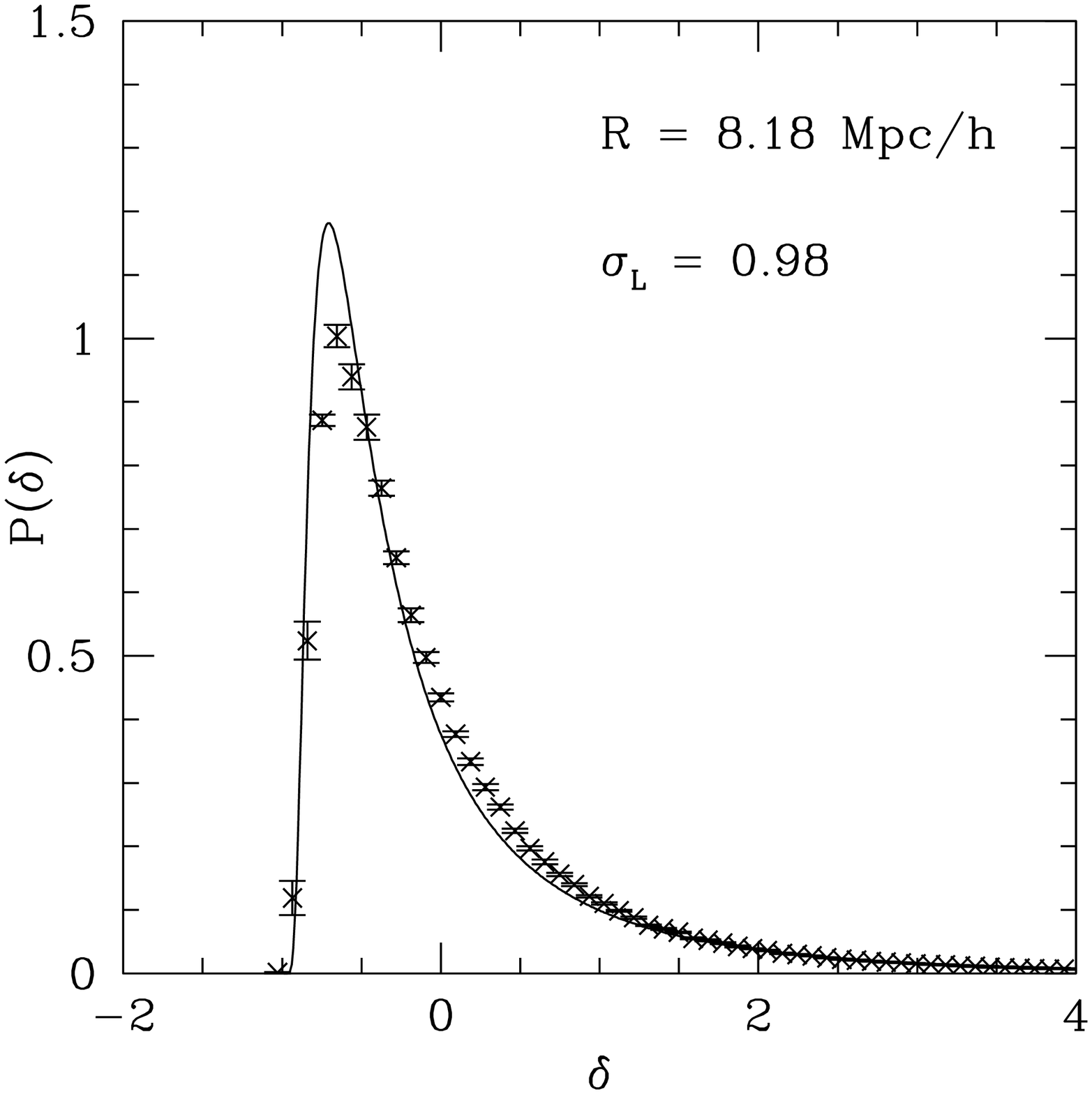}
\epsfxsize=5.7cm \epsfbox{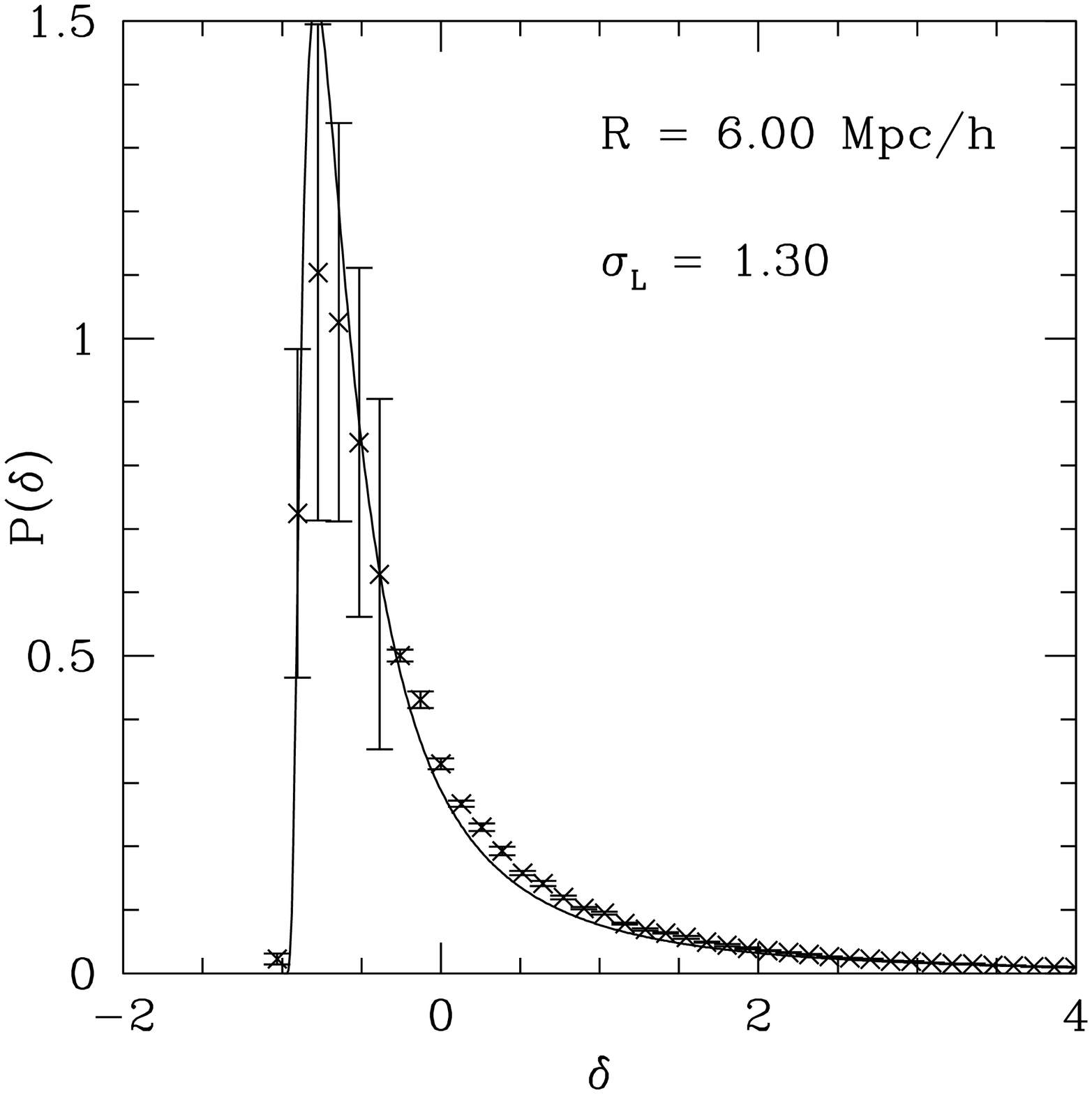}}
\centerline{\epsfxsize=5.7cm \epsfbox{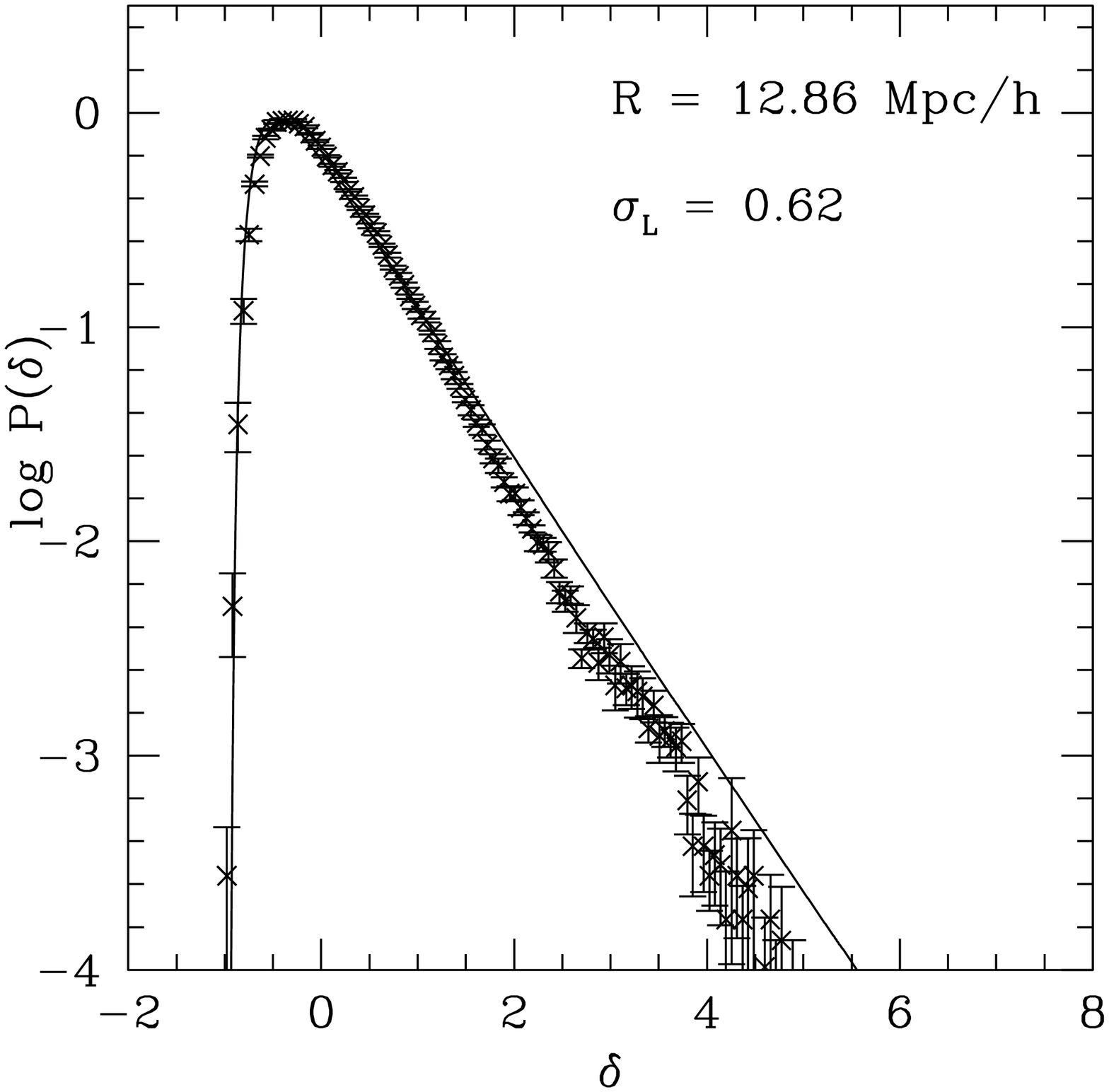}
\epsfxsize=5.7cm \epsfbox{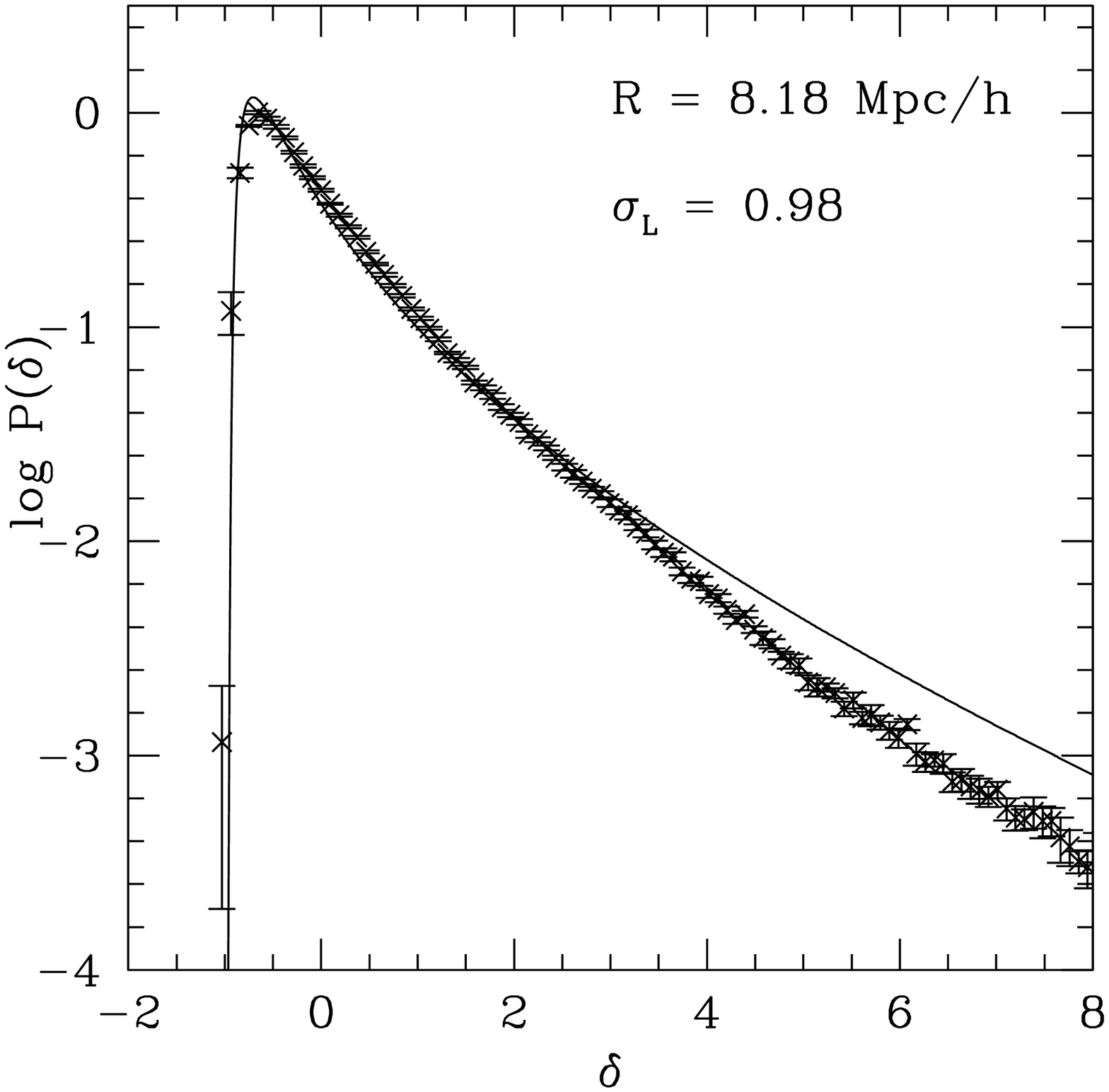}
\epsfxsize=5.7cm \epsfbox{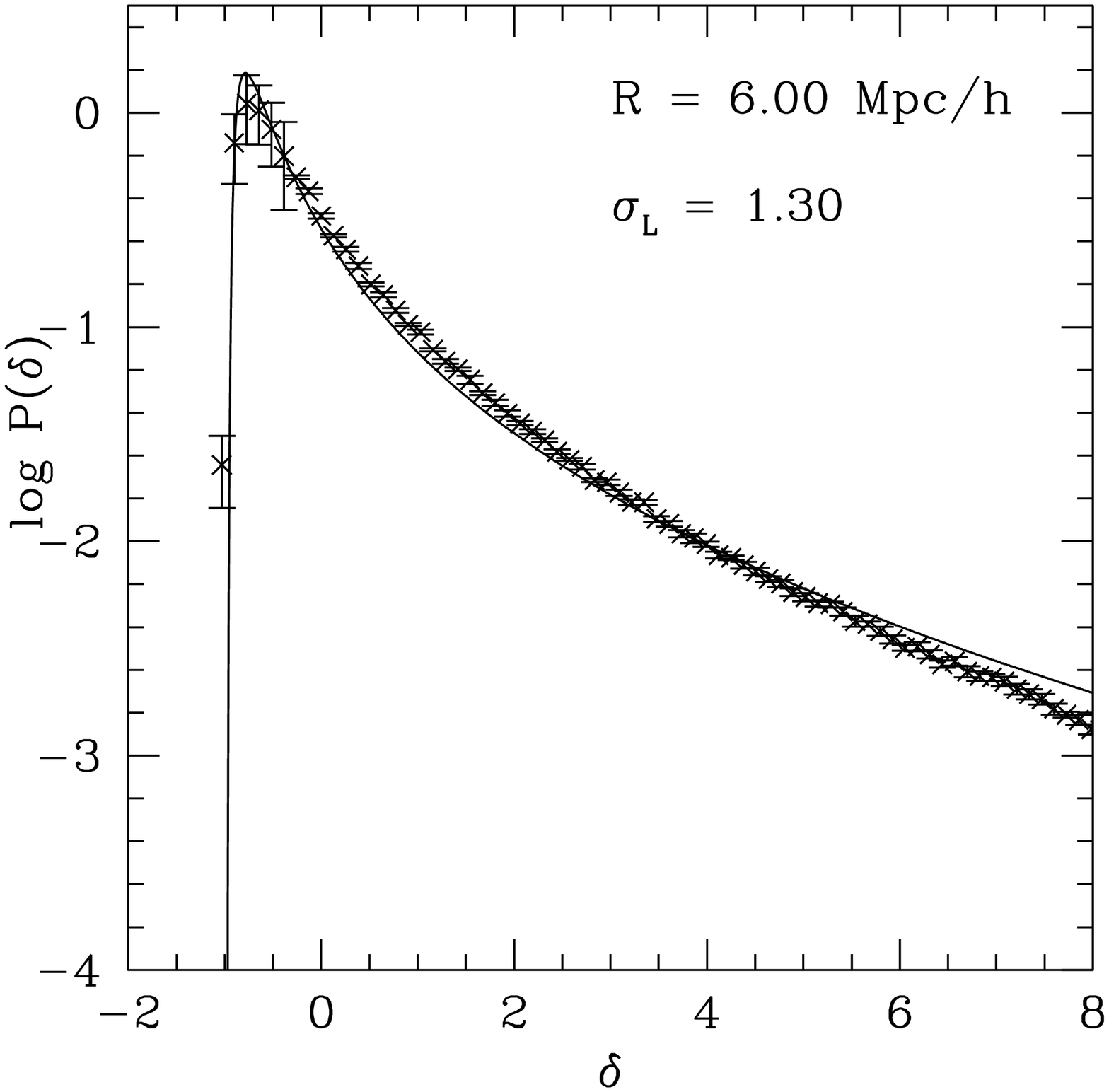}}
\caption[Fig1]{\label{fig1}The PDF $P[\delta]$ (top) and log $P[\delta]$
(bottom)
as a function of $\delta$ in the SCDM model.  Points
with $1-\sigma$ error bars are from the numerical simulation.  Solid curve
is the SC theoretical prediction. Each pair of panels corresponds
to a different smoothing scale $R$ and therefore a different 
linear rms fluctuation $\sigma_L$, indicated in the figure.}
\end{figure*}

\section{The PDF in the SC Model}

\subsection{Theory of the SC PDF}

One class of approximations to the evolution of large
scale structure are local Lagrangian mappings, in which
the density
at a Lagrangian point $\qvec$ at a time $t$ is taken to be a function
only of $t$ and the initial value of the density at $\qvec$,
\begin{equation}
\label{localdef}
\eta(\qvec,t) = N(t) f[\delta_L(\qvec)],
\end{equation}
where
$\eta = \rho /\bar \rho$, with $\bar \rho$ the
mean
density, and $\delta_L$ is the linearly-evolved
initial density fluctuation ($\delta_L=\eta_L-1$):
\begin{equation}
\delta_L(t) = D(t) \delta(t_0).
\end{equation}
In equation (\ref{localdef}), $N(t)$ is a normalizing factor
chosen so that $\int P[\eta] d\eta = 1$:
\begin{equation}
N(t)=\left< {1\over f(\delta_L)} \right>.
\label{norm1}
\end{equation}

A stochastic field of density fluctuations $\delta(\xvec)$ can be described
in terms of its cumulants $\kappa_p$, which are given by
$\kappa_2 \equiv \sigma^2 = \langle \delta^2 \rangle$, $\kappa_3 = \langle
\delta^3 \rangle$, $\kappa_4 = \langle \delta^4 \rangle
- \langle \delta^2 \rangle^2...$  An initially Gaussian
density field evolving gravitationally has the property
that $\kappa_p / \sigma^{2(p-1)}$ goes to a constant
in the limit where $\sigma$ goes to zero, where $\sigma$
is the rms density fluctuation (Peebles 1980,
Fry 1984, Bernardeau 1992).
Hence, it is conventional to define the hierarchical
amplitudes $S_p$ given by:
\begin{equation}
S_p(\sigma) \equiv {\kappa_p \over {(\sigma^2)^{p-1}}}
\end{equation}
The values of $S_p$ in the limit where $\sigma \rightarrow 0$
are called the tree-level hierarchical amplitudes.

This formalism can be exploited to derive an approximation
to the PDF in the quasi-linear regime.
For small fluctuations, a generic local Lagrangian
mapping can be expressed
in a Taylor series:
\begin{equation}
\label{Taylor}
\eta(\qvec) = f[\delta_L(\qvec)] = \sum_k {\nu_k\over{k!}} 
\delta^k_L(\qvec)
\end{equation}
It can be shown (PS, FG) that for the local Lagrangian
mapping given in equation (\ref{localdef}), expanded
as a Taylor series as in equation (\ref{Taylor}), the
hierarchical amplitudes are given by
\begin{eqnarray}
S_3(0) &=& 3 \nu_2, \\
S_4(0) &=& 4 \nu_3 + 12 \nu_2^2,
\end{eqnarray}
and so on.  These amplitudes can be made identical
to those of exact tree-level perturbation theory
if we take the mapping in Eq. (\ref{Taylor}) to be given
by (B92, FG):
\begin{eqnarray}
\label{SC1}
\eta &=& N{9 \over 2}{(\theta-\sin\theta)^2 \over
(1-\cos\theta)^3}, \nonumber\\
\delta_L &=& {3\over 5} [{3\over 4}(\theta-\sin\theta)]^{2/3},
\end{eqnarray}
for $\delta_L > 0$, $\eta > 1$, and
\begin{eqnarray}
\label{SC2}
\eta &=& N{9 \over 2}{(\sinh\theta-\theta)^2 \over
(\cosh\theta-1)^3}, \nonumber\\
\delta_L &=& - {3\over 5} [{3\over 4}(\sinh\theta-\theta)]^{2/3},
\end{eqnarray}
for $\delta_L < 0$, $\eta < 1$.
[In this paper, we assume Gaussian initial conditions, but
this result can be extended to non-Gaussian initial conditions
as well (Gazta\~naga \& Fosalba 1998)].

There is a physical basis for this result: 
in the limit where shear is neglected,
the equations of perturbation growth become (FG)
\begin{equation}
\label{SCdef}
\ddot \delta + {\dot a \over a} \dot \delta
- {4 \over 3} {\dot \delta^2 \over (1 + \delta)}
= 4 \pi G a ^2 \delta (1 + \delta),
\end{equation}
where the dot denotes the convective derivative with respect to conformal
time.  (Hence, equation (\ref{SCdef}) is an equation for the Lagrangian
density).  But equation (\ref{SCdef}) is identical to the equation for
$\delta$ in spherical collapse, so the solution is the local Lagrangian SC mapping given by
Eqs. (\ref{SC1})- (\ref{SC2}).  Furthermore, the Fourier-space integrals which
give the tree-level hierarchical amplitudes are spherically symmetric,
guaranteeing that the SC approximation gives the correct hierarchical
amplitudes at tree level (FG).  Although this approximation was dubbed the
``spherical-collapse approximation" (FG), it does not presume or imply that the
actual physical collapse is spherically symmetric; rather, it proceeds from
the approximation that the clustering is shear-free.  This approximation
should not be confused with the well-known analytic results for actual
spherical collapse (Peebles 1980).

For the SC approximation we have:
\begin{eqnarray}
\nu_2 &=& {34\over 21} \sim 1.62, \nn \\ 
\nu_3 &=& {682\over 189} \sim 3.61, \nn \\ 
\nu_4 &=& {446440\over 43659} \sim 10.22, \nn \\
\nu_5 &=& {8546480\over 243243} \sim 35.13,
\label{nusc}
\end{eqnarray}
and note that $\nu_1=1$ to produce the linear theory result.
As noted, these numbers give the correct contribution to
the $p$th-order cumulants, $\kappa_p$, for Gaussian initial conditions,
e.g., $S_3=34/7$ (Peebles 1980, Section 42).

We now derive the PDF in the SC approximation.
The SC local Lagrangian mapping which relates $\delta_L(\qvec)$
to $\eta(\qvec)$ is given by
Eqs. (\ref{SC1}) and (\ref{SC2}).
In our discussions below, the mapping $\eta = Nf(\delta_L)$
is always understood to refer to equations (\ref{SC1}) and (\ref{SC2}),
although one cannot write down a non-parametric form for $f(\delta_L)$.
These equations also define an inverse function:  $\delta_L = f^{-1}(\eta/N)$;
it is $\delta_L(\eta/N)$ as
defined by equations (\ref{SC1}) and (\ref{SC2})
which actually enters into our
equations for the PDF.
The evolved PDF (given, as usual, in Eulerian space)
is (PS)
\begin{equation}
\label{PDF1}
P[\eta] = P_0\left[\delta_L(\eta/N)\right]{1 \over \eta} {d\delta_L(\eta/N) \over d\eta}
\end{equation}
where $P_0$ is the initial (unevolved) density distribution function.
The normalizing factor $N$ is most easily derived from:
\begin{equation}
\label{N1}
N = \int_0^\infty 
P_0(\delta_L(x)) {1 \over x} {d\delta_L(x) \over dx} dx,
\end{equation}
where we have used the change of variables $x = \eta/N$.
For Gaussian initial conditions, $P_0$ is a Gaussian, and we get
\begin{equation}
\label{PDF2}
P[\eta] = {1\over \sqrt{2\pi} \sigma_L}
\exp\left[-{\delta_L(\eta/N)^2\over{2\sigma_L^2}}\right]{1 \over \eta} {d\delta_L(\eta/N) \over d\eta}
\end{equation}
where
$\sigma_L$ is the linearly evolved rms fluctuation:
\begin{equation}
\sigma_L(t) = D(t) \sigma_0,
\end{equation}
and $N$ is given by
\begin{equation}
\label{N2}
N = \int_0^\infty {1\over \sqrt{2\pi} \sigma_L}
\exp\left[-{\delta_L(x)^2\over{2\sigma_L^2}}\right]
{1 \over x} {d\delta_L(x) \over dx} dx.
\end{equation}

This expression gives the PDF for an unsmoothed density
field, which is not physically observable.  If we smooth
the final density field with a spherical top-hat window function,
then we obtain a new ``smoothed" local Lagrangian mapping density
$f_S(\delta_L)$, which is given in terms of the unsmoothed
mapping $f$ through the implicit relation
(Bernardeau 1994; PS; FG):
\begin{equation}
\label{smooth1}
f_S(\delta_L) = f\biggl(\delta_L {\sigma(f_S(\delta_L) R_0) \over \sigma(R_0)}\biggr)
\end{equation}
where $R_0$ is the radius of the spherical top-hat,
and $\sigma(R_0)$ is the rms fluctuation of the linear
density field (in practice, derived by integrating over
the linear power spectrum).  Taking $\eta_S = N f_S(\delta_L)$,
where $N$ is the new (smoothed) normalization factor,
this expression can be
recast in the form
\begin{equation}
\label{smooth2}
\delta_L(\eta_S/N) = f^{-1}(\eta_S/N){\sigma(R_0) \over \sigma(R_0 \eta_S/N)}.
\end{equation}
If the linear power spectrum
is a power law, $P(k) \propto k^n$, equations
(\ref{smooth1}) and (\ref{smooth2}) simplify to (Bernardeau 1994)

\begin{equation}
\label{smooth3}
f_S(\delta_L)=f\left[ \delta_L f_S(\delta_L)^{-{(n+3)}/6 }\right],
\end{equation}
which gives
\begin{equation}
\label{smooth4}
\delta_L(\eta_S/N) = f^{-1}(\eta_S/N) [\eta_S/N]^{(n+3)/6}
\end{equation}
The expression for $\delta_L$ from either equation (\ref{smooth2}) or equation
(\ref{smooth4}) 
can be substituted
into equations (\ref{PDF2})-(\ref{N2}) to
obtain $P_S(\eta)$ (the PDF of the smoothed density field).

\subsection{Test with Nbody simulations}
\label{sec:nbody}

To test our approximations for the PDF, we compare to a set of 10
standard cold dark matter (SCDM) simulations with $64^3$ particles
over an $L = 180$ Mpc/$h$ box.  (We also used simulations with an 
$L \simeq 400$ Mpc/$h$ box to test for volume effects.
See Baugh, Gazta\~naga \& Efstathiou 1995, for more details on 
these simulations).  By smoothing over different lengths
we derived a set of PDF's for different values of the linear
rms fluctuation $\sigma_L$.  

The results are displayed in Fig. \ref{fig1}.
It is clear that the SC approximation provides excellent agreement
with the numerical PDF even for $\sigma_L \simgt 1$, except for
the large-$\delta$ tail of the distribution.  These
results are consistent with those of
Protogeros, Melott, \& Scherrer (1997), who
used a less-accurate
Lagrangian mapping which mimics the SC approximation
and found good agreement with the PDF for power-law initial
conditions.

\section{The Redshift Distortion of the PDF}

\subsection{Preliminaries}

A particle at a position $\rvec$ in real space, with local velocity
$\vvec$, will be measured to lie at a position $\rvec_{(z)}$
in redshift space, given by:
\begin{equation}
\rvec_{(z)} = \rvec + (v_r/H) \hat \rvec
\end{equation}
where $H$ is the Hubble parameter, and
$\hat \rvec$ points radially outward from the observer.
For simplicity, we use the ``infinite observer" approximation,
so that the redshift direction is taken to lie
parallel to one of the Cartesian axes, which
we take to be the $x$ axis.  Then we have:
\begin{equation}
\rvec_{(z)} = \rvec + (v_x/H) \hat \xvec.
\end{equation}
It is convenient to make the substitution
$\uvec = \vvec/H$, so that
\begin{equation}
\rvec_{(z)} = \rvec + u_x \hat {\bf x}.
\end{equation}
From mass conservation, we obtain the relation between
$\eta_{(z)}$, the value of $\eta$ measured at a particular
point in redshift space, and the corresponding
$\eta$ at the same (Lagrangian) point in real space,
where we can treat the redshift-space distortion
as a Lagrangian mapping:
\begin{equation}
\label{redshift}
\eta_{(z)} = {\eta \over 1 + (du_x/dx)},
\end{equation}
(see also Hui et al. 2000).
Equation (\ref{redshift}) is exact (for the infinite-observer case)
and can be used, in principle, to calculate the redshift-distortion
of the PDF.  Such a calculation can be carried out
for the skewness (Bouchet et al. 1995; Hivon et al. 1995;
Scoccimarro et al. 1999), but there is no simple way to apply
it to the full PDF.

Hence, we make an approximation
which is appropriate to the SC model.  First we take
$du_x/dx$ to be given by
\begin{equation}
{du_x \over dx} = {1\over 3} \theta,
\end{equation}
where $\theta \equiv \nabla \cdot \uvec$.
This relation between $du_x/dx$ and $\theta$ is
what would be expected in the limit
where shear (and vorticity) are neglected, and our approximation
has
the advantage that the statistical properties
of $\theta$
are well-known in tree-level perturbation theory.
We get
\begin{equation}
\label{redSC1}
\eta_{(z)} = {\eta \over 1 + (1/3) \theta}
\end{equation}
which gives, for $\delta$,
\begin{equation}
\label{redSC2}
\delta_{(z)} = {\delta - (1/3) \theta \over 1 + (1/3) \theta}
\end{equation}

We can use the continuity equation:
\begin{equation}
{d\eta\over{dt}} +\,H\,\eta \, \theta =0,
\end{equation}
to express $\theta$ as a Lagrangian mapping of $\delta_L$:
\begin{equation}
\theta= - {1\over{H\,\eta}} \, {d\eta\over{dt}}
= -f_\Omega \,\, {\delta_L\over{\eta}} {d\eta\over{d\delta_L}},
\label{theta2}
\end{equation}
where $f_\Omega \equiv d \ln D/d \ln a$, comes from applying the
chain rule to the derivative
of the linear growth factor: $\delta_L = D(t) \delta_0$.

\subsection{Hierarchical moments}

Consider first the case of an unsmoothed density field.
For small fluctuations all we need are the coefficients of the
corresponding Taylor expansion:

\begin{equation}
\eta_{(z)} = \sum_k {\nu_{k(z)}\over{k!}}~\delta^k_L.
\end{equation}
We find for the first few orders:
\begin{eqnarray}
\nu_{1(z)} &=& 1+ \epsilon \\ \nn
\nu_{2(z)} &=& \nu_2 + 2 \epsilon \nu_2 + 2 \epsilon^2\\ \nn
\nu_{3(z)} &=& \nu_3 + 3 \epsilon \nu_3 + 6 \epsilon^2 (2\nu_2-1) + 6 \epsilon^3\\
\nn
\nu_{4(z)} &=& \nu_4 + 4 \epsilon \nu_4 + 24 \epsilon^2 (\nu_3+\nu_2^2
-5\nu_2/2+1) \\ \nn &+& 24 \epsilon^3 (3\nu_2-2) +\epsilon^4 \nn,
\end{eqnarray}
where $\epsilon \equiv f_\Omega/3$.
Note that in redshift space
$\nu_{1(z)} \neq 1$, because of the ``Kaiser'' effect.

\begin{figure}
\centering\centerline{
{\epsfxsize=7cm \epsfbox{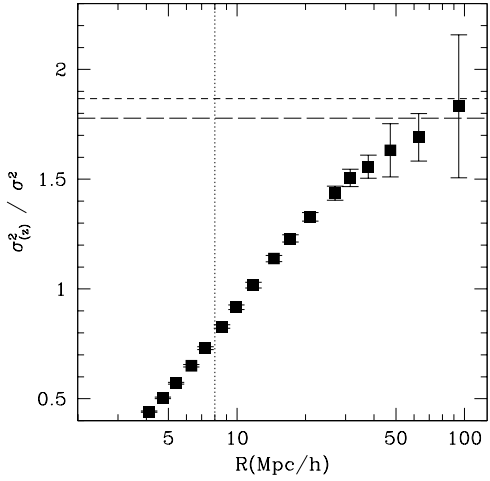}}}
\caption[]{\label{beta189} The ratio of the redshift to
real space variance $ \sigma^2_{(z)}/\sigma^2$ as a function of
smoothing scale in SCDM simulations
(points with error-bars) as compared to the ``Kaiser" results (short-dashed
lines) and the linear SC model (long-dashed lines). 
The dotted vertical line marks
the scale where the linear variance is unity, i.e. $\sigma_L=1$.  Error
bars are $1-\sigma$.}
\end{figure}

First consider the hierarchical amplitudes.
For the redshift variance we have:
\beq
\sigma^2_{(z)} = (1+\epsilon)^2 \sigma_L^2 
+ \Or \left( \sigma_L^4 \right).
\label{sigmaz}
\eeq
Note that this gives a reasonable approximation to the ``Kaiser" effect
in Eq. (\ref{Eq:Kaiser}). For example, for
 $f_\Omega \rightarrow 1$ ($\epsilon
\rightarrow 1/3$), the expression above gives
 $\sigma^2_{(z)}/\sigma^2_L = 16/9 \simeq 1.78$, compared
with  $\sigma^2_{(z)}/\sigma^2_L = 28/15 \simeq 1.87$ 
in Eq. (\ref{Eq:Kaiser}). These results are compared with
(fully non-linear) SCDM simulations in Fig. \ref{beta189}.

For the redshift-space skewness $S_{3(z)}$ and kurtosis $S_{4(z)}$,
at leading order, are:
\bea
\label{S3(z)}
S_{3(z)} &=&  {3 \nu_{2(z)}\over{\nu_{1(z)}^2}} = {{S_3 + 2 \epsilon S_3 + 6
\epsilon^2}\over{(1+\epsilon)^2}} + \Or \left( \sigma_L^2 \right), \\ \nn
S_{4(z)} &=& {{S_4+ 4 \epsilon S_4 + 3 \epsilon^2 (S_4+4 S_3^2/9+32 S_3/3 -8)
}\over{(1+\epsilon)^4}} \\ \nn &+& {48 \epsilon^3 S_3 + 72 \epsilon^4
\over{(1+\epsilon)^4}} + \Or \left( \sigma_L^2 \right) , 
\eea
where $S_3=3 \nu_2$ and $S_4= 12 \nu_2^2+4\nu_3$ are the corresponding 
real-space leading order results.
In the limit $f_\Omega \rightarrow 0$ we get the real-space results, as expected.
In the limit $f_\Omega \rightarrow 1$ ($\epsilon \rightarrow 1/3$) we have,
at leading order:
\bea
S_{3(z)} &=& {15\over{16}} S_3 + {3\over{8}} \label{s3zs4z}, \\ \nn 
S_{4(z)} &=& {27\over{32}} S_4 + {3\over{64}} S_3^2 + {27\over{16}} S_3
-{9\over{16}}.
\eea
So redshift distortions are not very large at leading order even in the
case when $f_\Omega \rightarrow 1$.
For the skewness, we get $S_{3(z)} = 4.93$, in excellent agreement with
the results of exact perturbation theory, which predicts
$S_{3(z)} = 5.03$ (Hivon et al. 1995), a difference of roughly
$2\%$.  For the kurtosis, we obtain
$S_{4(z)} = 47.5$, (no exact calculation of $S_{4(z)}$ has been done
thus far), which compares with a real-space value of
$S_4 = 45.9$.  Like the skewness, the kurtosis is only
slightly increased by redshift distortions.

In practice, one needs to reach very large scales
($R \simgt 100 ~h^{-1} $Mpc) to
get to the regime where leading order is a good approximation.
This is illustrated in Fig. \ref{beta189}, which shows how the ratio
$\sigma^2_{(z)}/\sigma^2$ estimated from SCDM Nbody simulations
(presented in Section \ref{sec:nbody}), only reaches
the linear regime at  $R \simgt 100 ~h^{-1} $Mpc, even when the simulated
box is $L= 400 ~h^{-1} $Mpc and the scale of non-linearity is $8 ~h^{-1} $Mpc.

We should therefore consider next to leading order corrections.
For the variance we have:
\bea
\sigma_{(z)}^2 &=& s_{2,2(z)} \sigma_L^2 + s_{2,4(z)} \sigma_L^4, \\ \nn
 s_{2,2(z)} &=& (1+\epsilon)^2, \\ \nn
s_{2,4(z)} &=&  {8\over{3}}s_{2,4} + {212 S_3^2 +S_3 -716 \over{162}}.
\eea
For the skewness
and $f_\Omega \rightarrow 1$:
\bea
S_{3(z)} &=& S_{3,0(z)} +  S_{3,2(z)} \sigma_L^2, \\ \nn
S_{3,0(z)} &=& {15\over{16}} S_{3,0} + {3\over{8}}, \\ \nn
S_{3,2(z)} &=& {28\over{9}} S_{3,2} \\ \nn &+& 
{3772 - 1608 S_3 - 1149 S_3^2  - 13 S_3^3  + 864 S_4
\over{486}},
\eea
where the values on the right-hand side are the ones in real space.
We can see from the expressions above that the corrections 
are quite different in real and redshift space.

Now consider the effects of smoothing. For power-law
initial conditions, with a power spectrum $P[k] \propto k^n$, it
is possible to show that (Juszkievicz, Bouchet
and Colombi 1993)
\begin{equation}
S_{3S} = S_3 + \gamma,
\end{equation}
where
$\gamma = -(n+3)$.  Bernardeau (1994) has shown that the
above result is also true for a generic power spectrum, where
$\gamma=\gamma(R) = d \log(\sigma^2)/d \log(R)$.
In Section 3.4 below, we will argue
that the effects of smoothing on the redshift-distorted
PDF are identical to the effects in real space (e.g.,
Eqs. \ref{smoothz1}-\ref{smoothz2}).  Hence, we expect that
\begin{equation}
S_{3(z)S} = S_{3(z)} + \gamma
\end{equation}
Substituting these two equations into Eq. (\ref{S3(z)}),
we obtain
\begin{equation}
S_{3(z)S} = {S_{3S} + 2 \epsilon S_{3S} + 6 \epsilon^2
+ \gamma \epsilon^2) \over (1 + \epsilon)^2}.
\end{equation}
In the limit where $f_\Omega \rightarrow 1$, this becomes
\begin{equation}
S_{3(z)S} = {15\over 16}S_{3S} + {3\over 8} + {\gamma\over 16}.
\end{equation}
Since $\gamma/16$ is generally small compared to the other terms
in the equation, smoothing has a minimal effect
on the relation between the real and redshift space skewness.

\begin{figure}
\centering\centerline{
{\epsfxsize=7cm \epsfbox{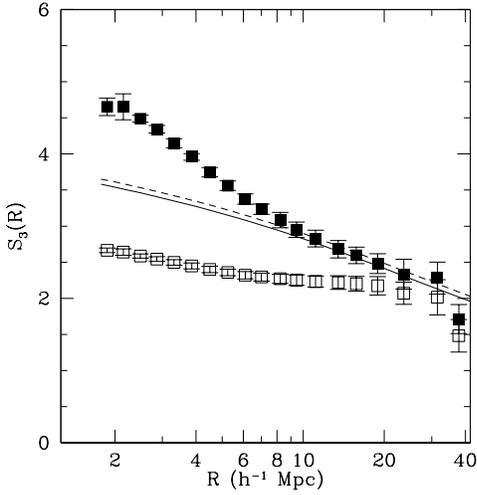}}}
\caption[fig]{\label{cics3c189z} Comparison of the redshift (open squares)
  and real (closed square) space skewness $S_3$, with
  $1-\sigma$ error bars, as a function of smoothing
  scale $R$, in SCDM simulations. Solid and dashed lines show the
  leading order SC result in real and redshift space.}
\end{figure}

Fig. \ref{cics3c189z} shows a comparison of the leading order contribution
to $S_3$ in redshift (dashed line) and real (solid line) space as 
a function of smoothing scale $R$ (which enters in the above expressions
through $\gamma=\gamma(R)$). As can be seen in the figure, the real
space predictions work well for $R \ge 10$, while in redshift space
we need to go to larger scales to find a similar agreement with 
leading order results.  (This is similar to what happens for the variance,
e.g., see Fig. \ref{beta189}).

\subsection{Dimensionless moments}

In discussions of large-scale structure, one is usually interested
in the hierarchical moments discussed in the previous section.
However, as we shall see in Section 3.5 below, we find a simple
empirical relationship between $P[\delta]$ and $P\left[\delta_{(z)}\right]$.
This result can be understood by considering the moments
of the normalized field
$\mu \equiv \delta/\sigma$.
For the variance we have:
\beq
\sigma_\mu^2 = \lexp \mu^2 \rexpc = 1,
\eeq
in both real and redshift space, by construction.
We can define
a dimensionless skewness:
\beq
B_3 \equiv \lexp \mu^3 \rexpc = S_3 ~\sigma,
\eeq
and in redshift space we find:
\bea
B_{3(z)} &=&    {{S_3 + 2 \epsilon S_3 + 6 \epsilon^2}\over{1+\epsilon}} \sigma_L
+ \Or \left( \sigma_L^3 \right),  \\ \nn
&=&   {(1 + 2 \epsilon)\over{1+\epsilon}}~B_3 + {{6
\epsilon^2}\over{1+\epsilon}} 
 ~\sigma_L + \Or \left( \sigma_L^3 \right),
\eea
which in the limit $f_\Omega \rightarrow 1$ gives:
\beq
B_{3(z)} = {5\over 4} B_3 + {1\over 2} \sigma_L + \Or \left( \sigma_L^3 \right),
\eeq
which should be compared
with the leading order real-space result $B_3= S_3 \sigma_L$.
At this order redshift distortions on the skewness are
therefore still relatively small, while by definition they are
zero in the variance. For the dimensionless kurtosis we find:
\beq
B_{4(z)} = {3\over 2}~B_4 + {1\over 12}~B_3^2 + (3 S_3-1) \sigma_L^2 + \Or \left(
\sigma_L^4 \right)
\eeq
The next to leading order term,
$B_{3,3} \sigma_L^3$  is, in the limit $f_\Omega \rightarrow 1$:
\bea
B_{3,3(z)} &\simeq& 2.95 -2.74 S_3 + 4.25 S_3^2 + 4.73 S_3^3 -1.6 S_4
\nn \\ &-& 5.8 S_3 S_4 + 1.24 S_5
\eea
while in real space ($f_\Omega \rightarrow 0$):
\bea
B_{3,3} &=& -2 + {{3}\over{2}}~S_3 +{{3}\over{2}}~S_3^2 + 
     {\frac{121}{108}}\,{{S_3}^3} + 
             {\frac{11}{8}} \,S_3~S_4 \nn \\&-& S_4 +
     {\frac{3}{10}}\,S_5
\eea
For the SCDM model at large scales, where
$\gamma \simeq -2$, we have: 
$S_3 \simeq 2.86$, $S_4 \simeq 13.9$ and $S_5 \simeq 97$.
We then find $B_{3,3(z)} \simeq 7.4$ while in real space:
$B_{3,3}  \simeq 1.16$. So there is a significant difference
in these higher corrections, although the resulting 
change in $B_3$ gets smaller as $\sigma_L \rightarrow 0$.

\begin{figure*}
\centering\centerline{
\epsfxsize=5.7cm \epsfbox{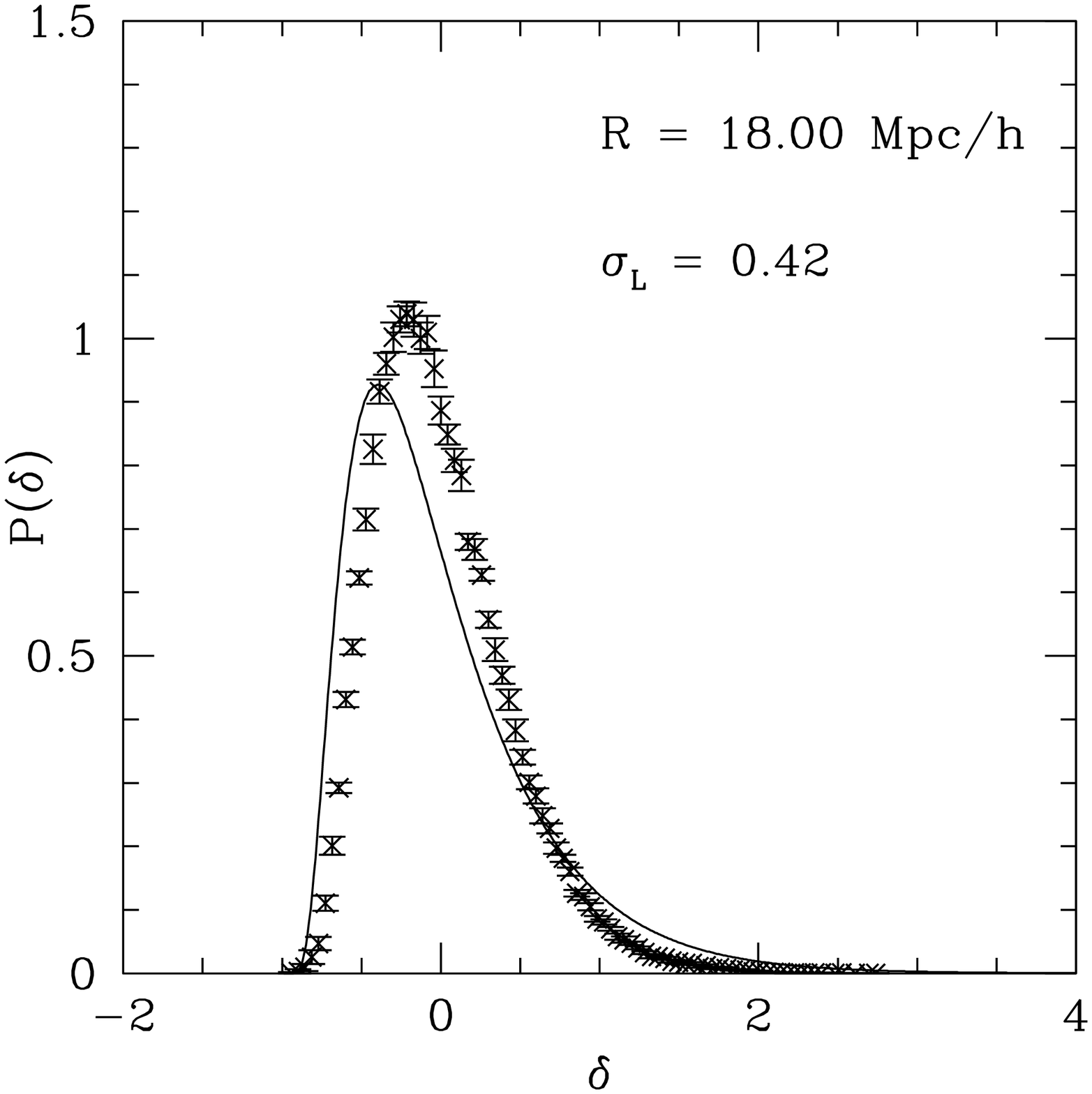}
\epsfxsize=5.7cm \epsfbox{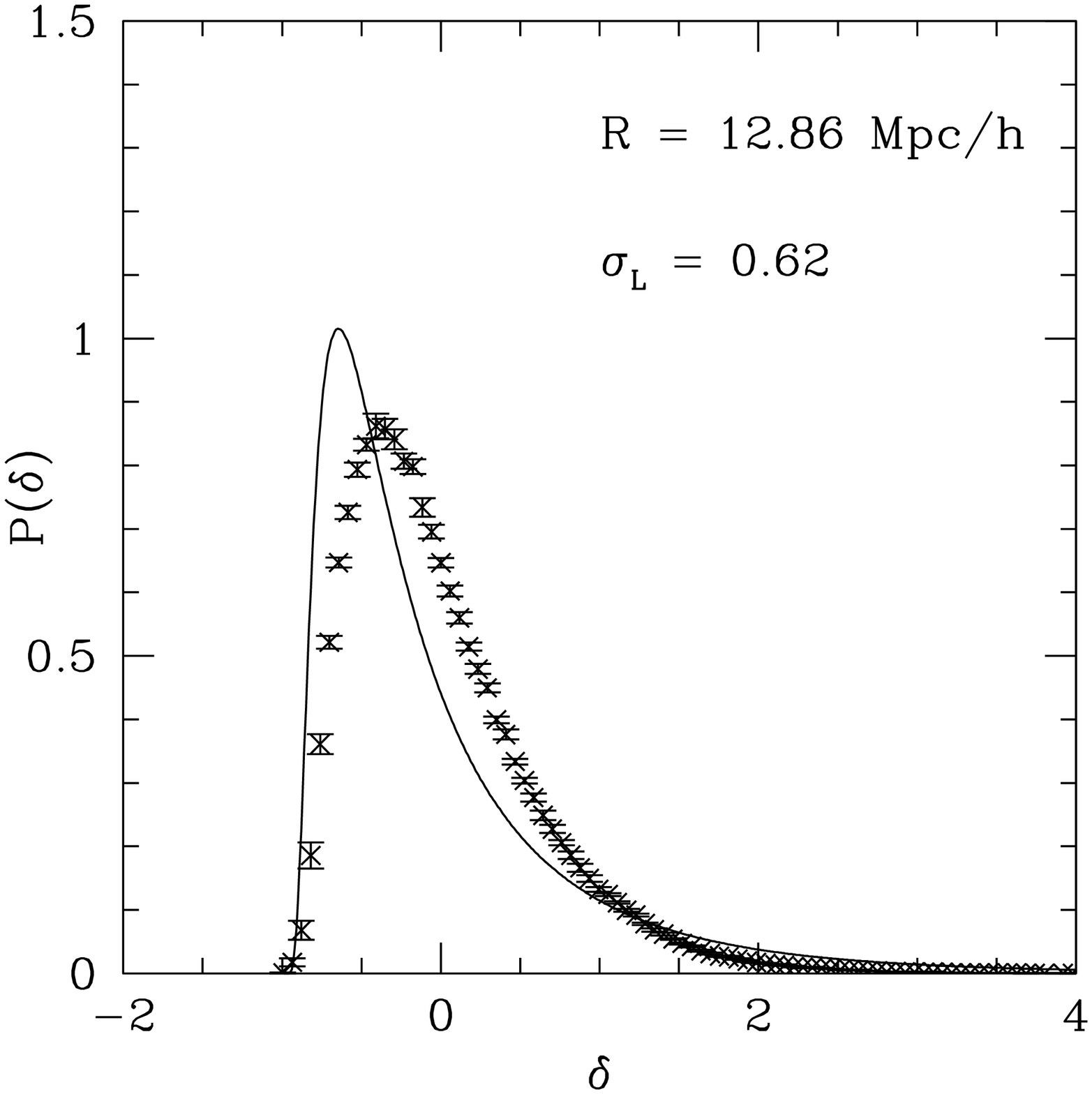}
\epsfxsize=5.7cm \epsfbox{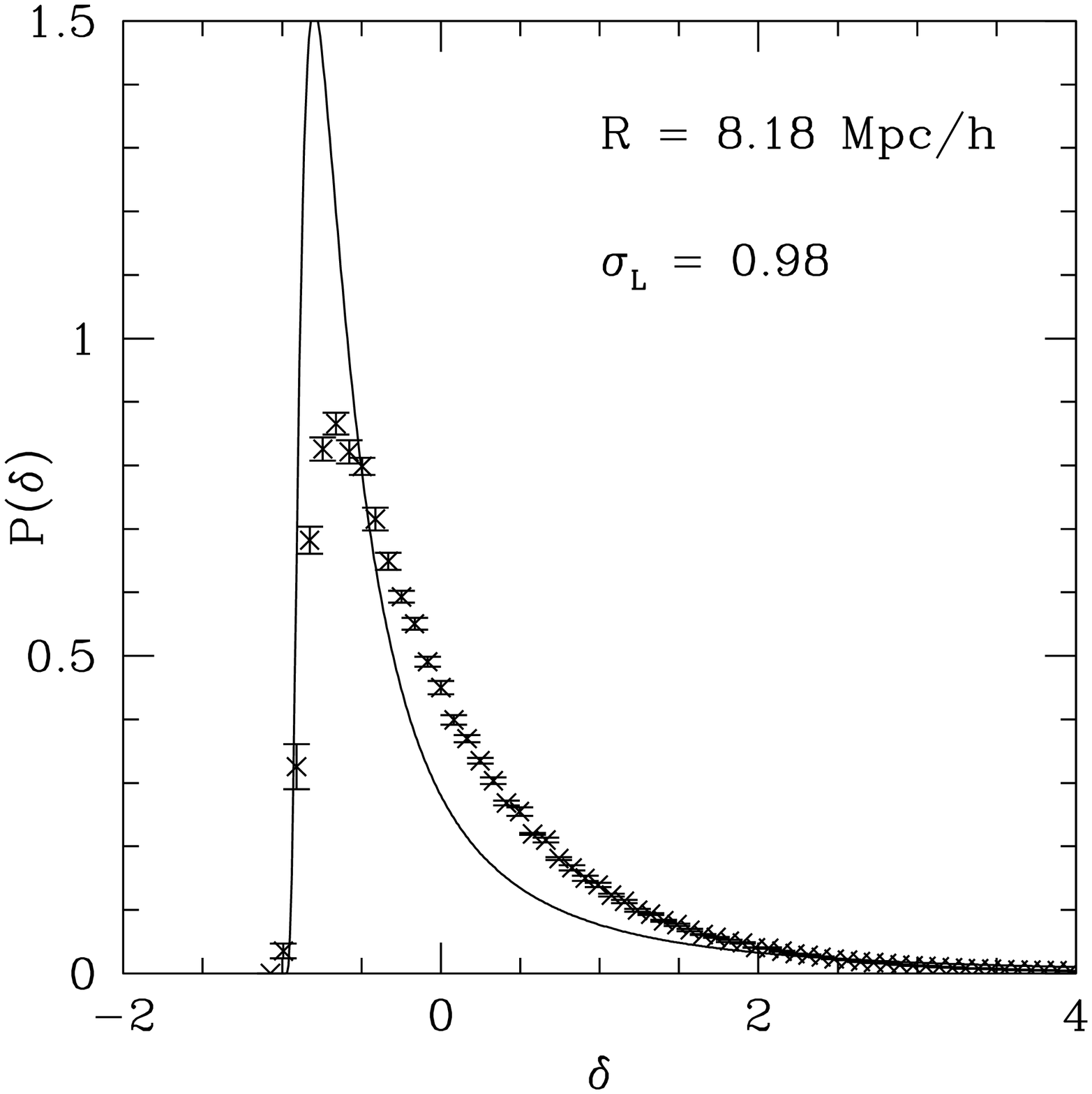}}
\centerline{
\epsfxsize=5.7cm \epsfbox{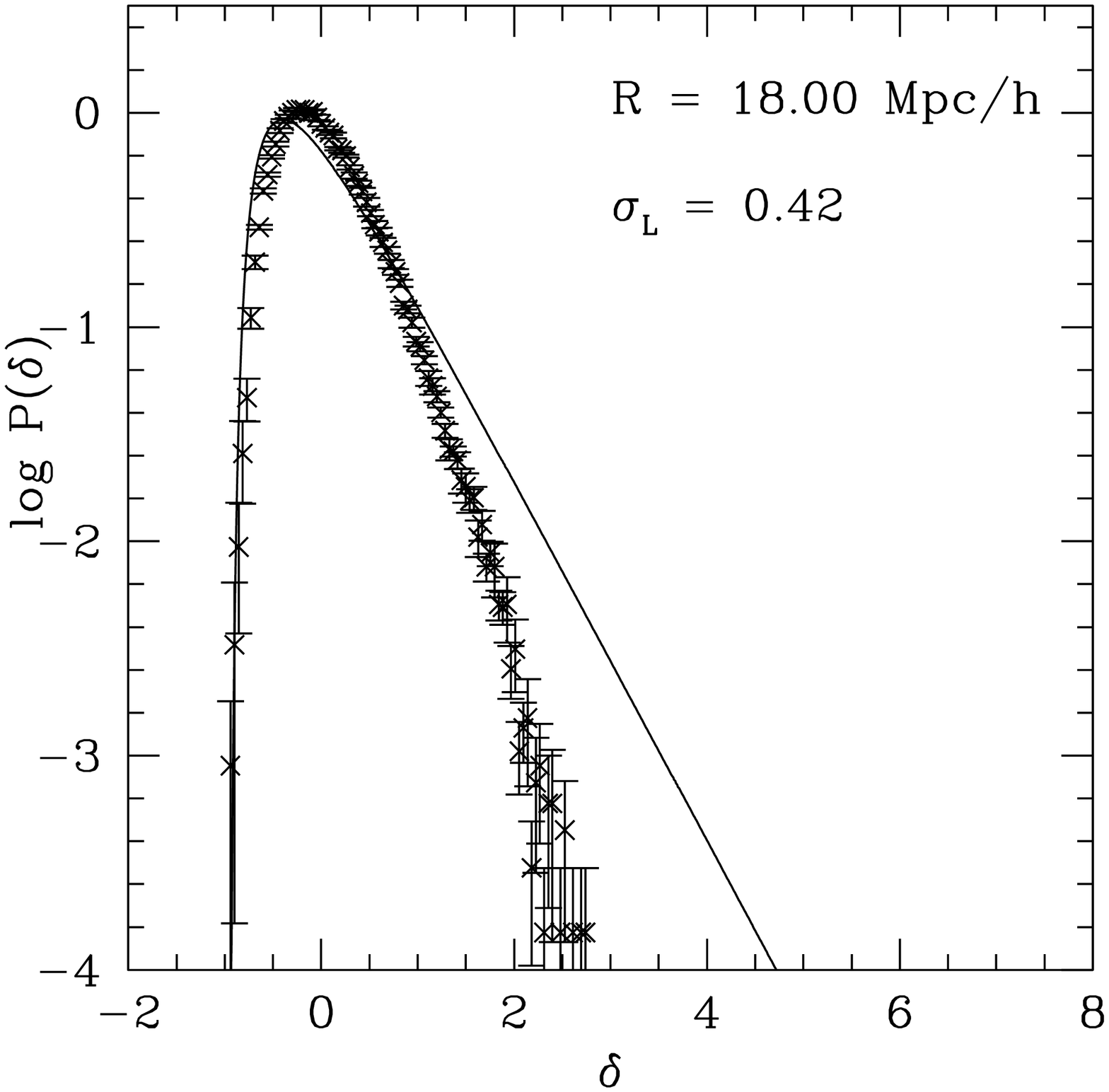}
\epsfxsize=5.7cm \epsfbox{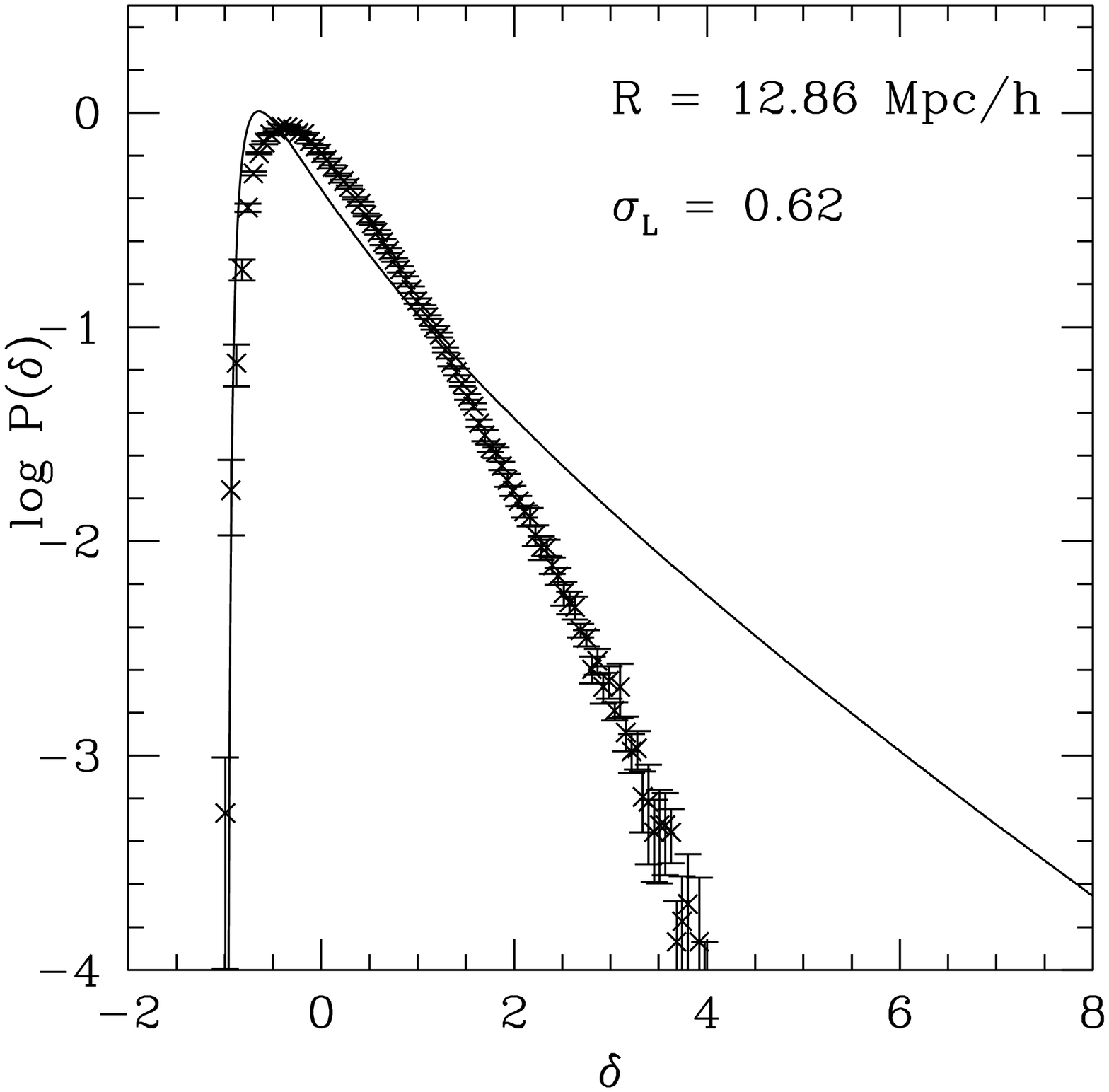}
\epsfxsize=5.7cm \epsfbox{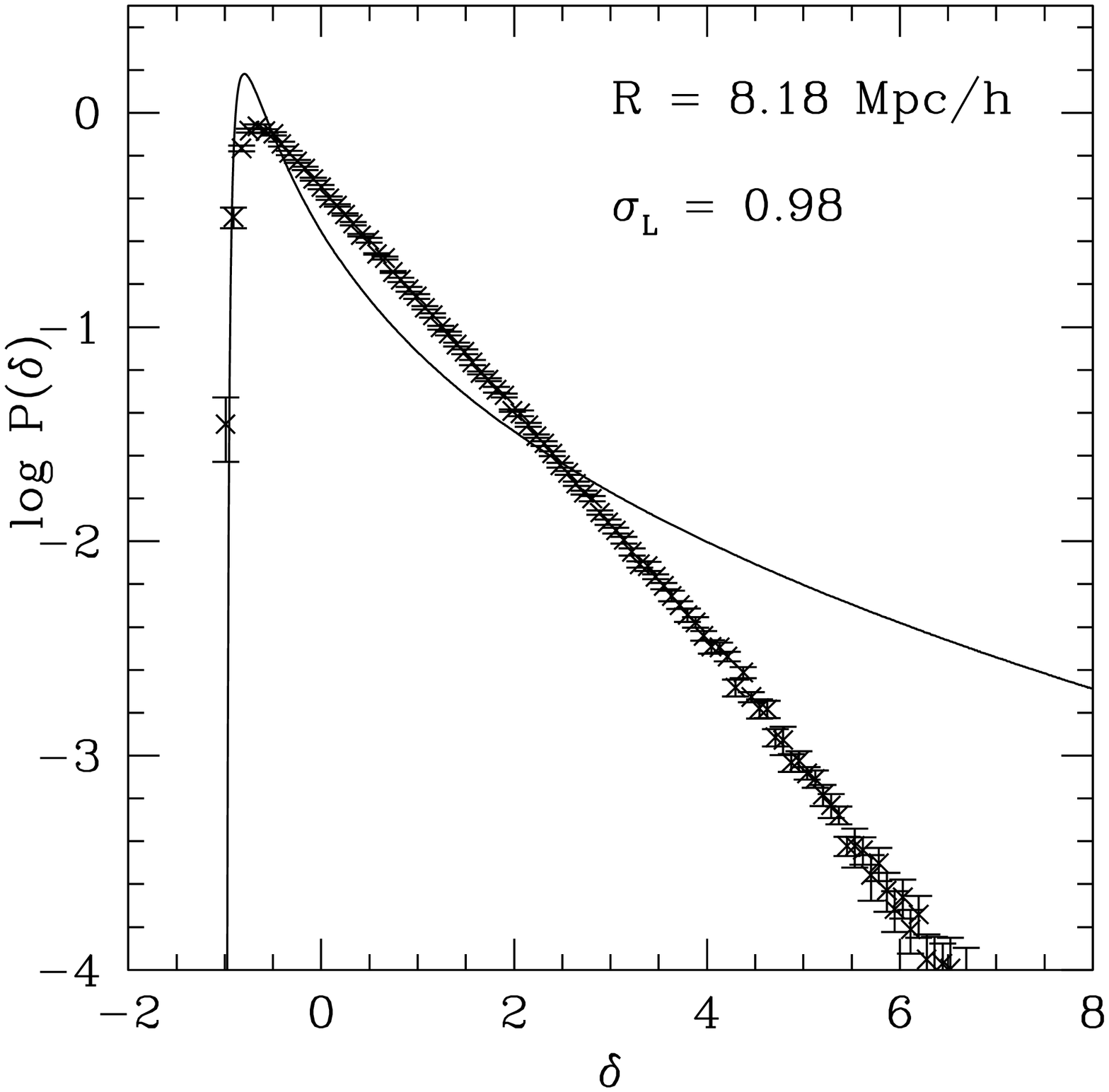}}
\caption[Fig2]{\label{fig2}The PDF $P[\delta]$ (top) and log $P[\delta]$
(bottom)
as a function of $\delta$ in redshift space.  Points
with $1-\sigma$ error bars are from the numerical simulation.  Solid curve
is the SC theoretical prediction. Each pair of panels corresponds
to a different smoothing scale $R$ and therefore a different linear rms
fluctuation $\sigma_L$, indicated in the figure.}
\end{figure*}

\subsection{The redshift-space PDF}

To obtain the final density at a given Lagrangian point, we first
apply the SC approximation to obtain $\eta$
as a function of $\delta_L$ in real space and then apply
the redshift mapping in
equation (\ref{redSC1}),
with $\theta$ given by equation (\ref{theta2}),
to go from $\eta$ to $\eta_{(z)}$.
The combination of these two mappings
can be treated as a single local Lagrangian mapping.
We obtain, for the local
Lagrangian relation between $\delta_L$ and
the redshift-distorted $\eta_{(z)}$:
\begin{equation}
\label{F}
\eta_{(z)} = F(\delta_L) \equiv {f(\delta_L) \over 1 - {1\over 3} f_\Omega
{\delta_L \over f({\delta_L})}
{df(\delta_L) \over d\delta_L}}
\end{equation}
where the function $f$ is given by equations (\ref{SC1}) and (\ref{SC2}).
Finally, we must include the effects of smoothing.  To do this, we
apply equation (\ref{smooth2}) or (\ref{smooth4}) to the redshift-distorted
density field.  For a power-law power spectrum, for example we obtain:
\begin{equation}
\label{smoothz1}
\delta_L(\eta_{(z)S}/N) = F^{-1}(\eta_{(z)S}/N) (\eta_{(z)S}/N)^{(n+3)/6},
\end{equation}
while in the general case, we get
\begin{equation}
\label{smoothz2}
\delta_L(\eta_{(z)S}/N) = F^{-1}(\eta_{(z)S}/N){\sigma(R_0)
\over \sigma(R_0
\eta_{(z)S}/N)}.
\end{equation}
In these equations, $F^{-1}$ is the inverse of the function
$F$, given in equation (\ref{F}), which transforms $\delta_L$ into $\eta_{(z)}$.

A couple of comments are in order regarding our smoothing methodology:
we have convolved the local Lagrangian mappings $\delta_L \rightarrow
\eta$ and $\eta \rightarrow \eta_{(z)}$ into a single local Lagrangian mapping
$\eta_{(z)} = F(\delta_L)$, given by equation (\ref{F}); it is this mapping
which enters into
Bernardeau's smoothing formalism.  Note also that
we are performing our operations in the correct order:
equations (\ref{F}) and (\ref{smoothz1})
or (\ref{smoothz2}) amount to first evolving the density field,
then applying the redshift distortion, and smoothing the final result.

For Gaussian initial conditions,
our final expression for the redshift-distorted PDF is then:
\begin{equation}
P\left[\eta_{(z)S}\right] = {1\over \sqrt{2\pi} \sigma_L}
\exp\left[-{\delta_L^2\over{2\sigma_L^2}}\right]{1 \over \eta_{(z)S}}
{d\delta_L \over d\eta_{(z)S}},
\end{equation}
where the function $\delta_L=\delta_L(\eta_{(z)S}/N)$ is given by equations
(\ref{F}) and (\ref{smoothz1}) or (\ref{smoothz2}), and $N$ is given by
\begin{equation}
N = \int_0^\infty {1\over \sqrt{2\pi} \sigma_L}
\exp\left[-{\delta_L(x)^2\over{2\sigma_L^2}}\right]
{1 \over x} {d\delta_L(x) \over dx} dx.
\end{equation}

We now test this approximation for the redshift-distorted PDF numerically.
The numerical simulations (described in Section \ref{sec:nbody}) were used to
calculate the redshift space PDF for a variety of values of $\sigma_L$.  These
are compared with the SC prediction for the redshift-space PDF in
Fig. \ref{fig2}.

It is clear that the SC approximation gives a reasonable fit
to the redshift-space PDF for $\sigma_L \simlt 0.4$
but fails for larger values.
Despite these problems, the SC approximation appears applicable
to larger values of $\sigma_L$ than does the
the Kaiser expression
for $\sigma_{(z)}$  (Eq. \ref{kaiser}).  For example, Eq. (\ref{kaiser}) predicts
$\sigma_{(z)}^2/\sigma^2 = 1.87$.  In our numerical simulations,
we obtain
$\sigma_{(z)}^2/\sigma^2 = 1.21$ for $\sigma_L = 0.42$,
corresponding to the left panel in Fig. \ref{fig2}.  
Thus, Eq. (\ref{kaiser}) fails badly
for $\sigma_L$ as small as 0.4 (see also Fig. \ref{beta189}), 
for which the SC approximation
provides reasonable agreement with the empirical PDF.

\subsection{An interesting empirical result}

\begin{figure*}
\centering\centerline{
\epsfxsize=5.7cm \epsfbox{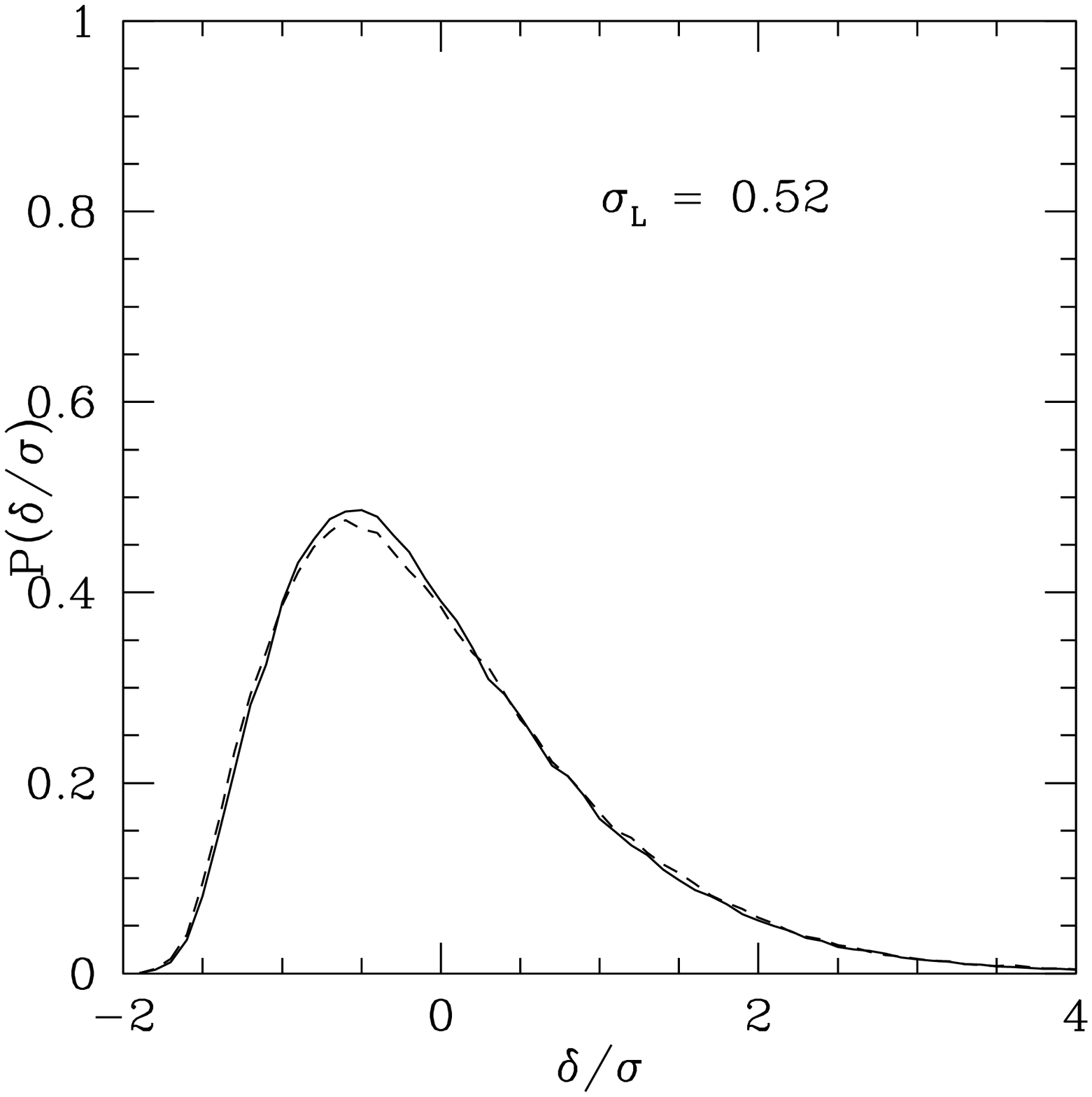}
\epsfxsize=5.7cm \epsfbox{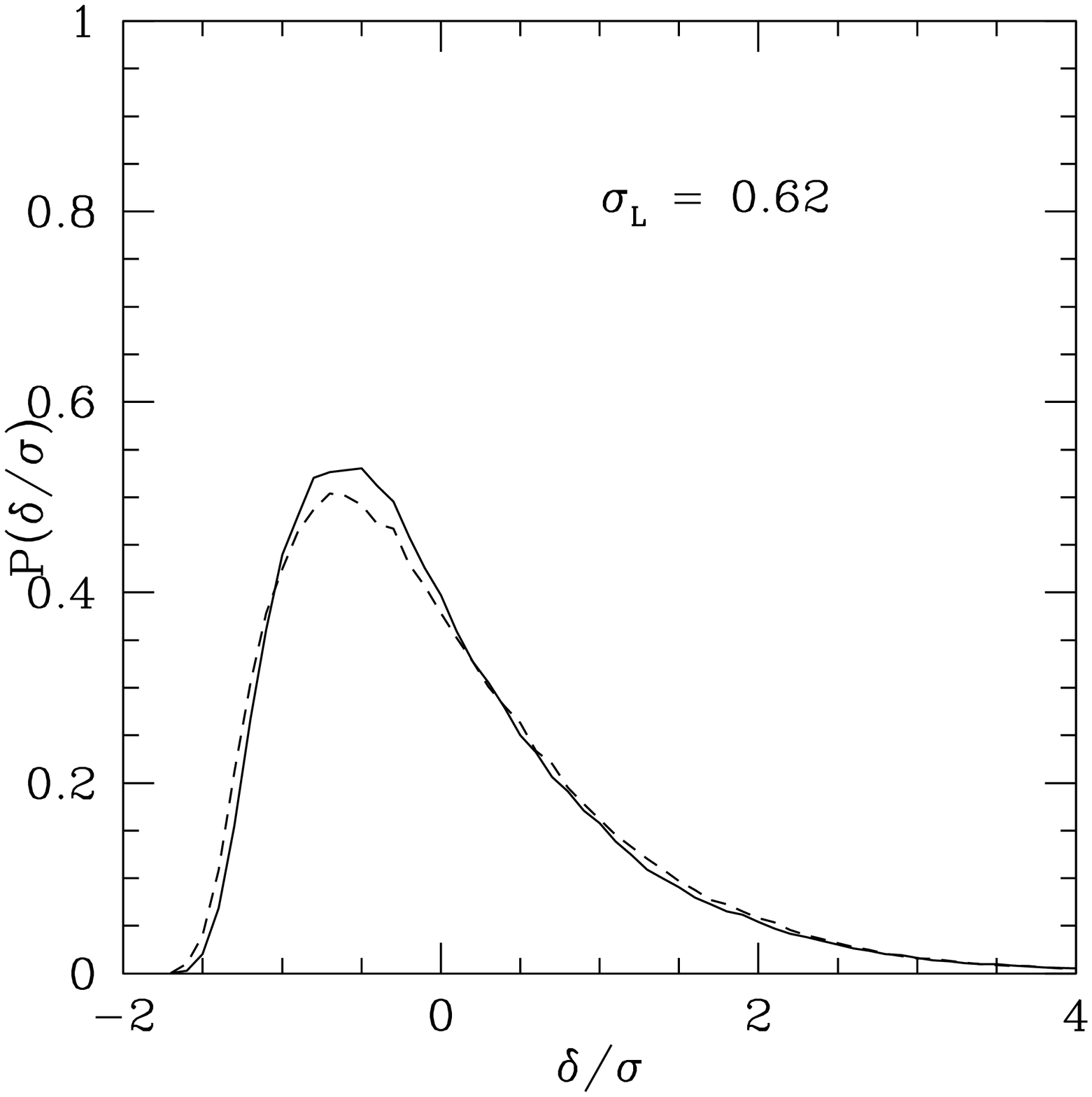}
\epsfxsize=5.7cm \epsfbox{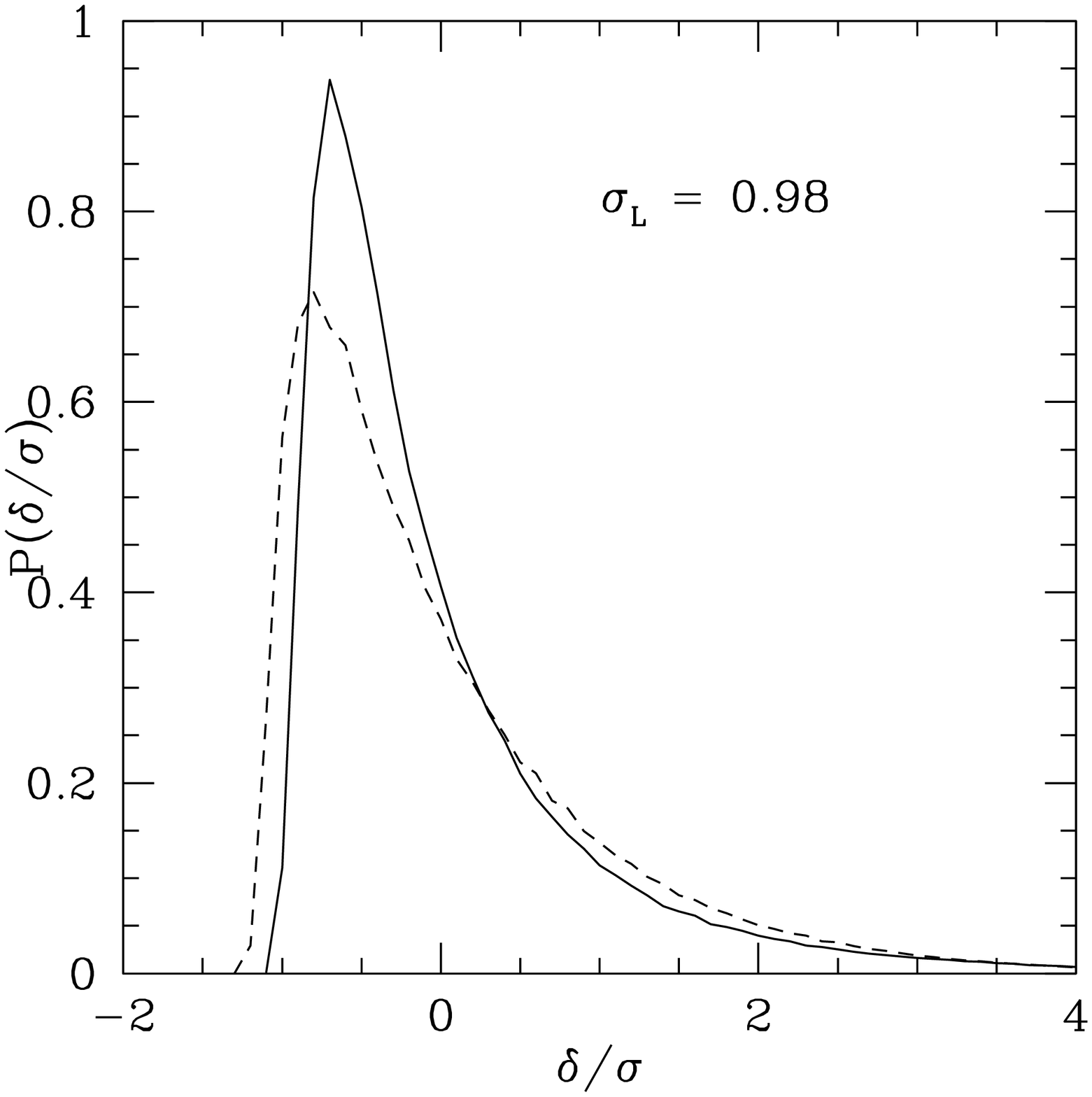}}
\centerline{\epsfxsize=5.7cm \epsfbox{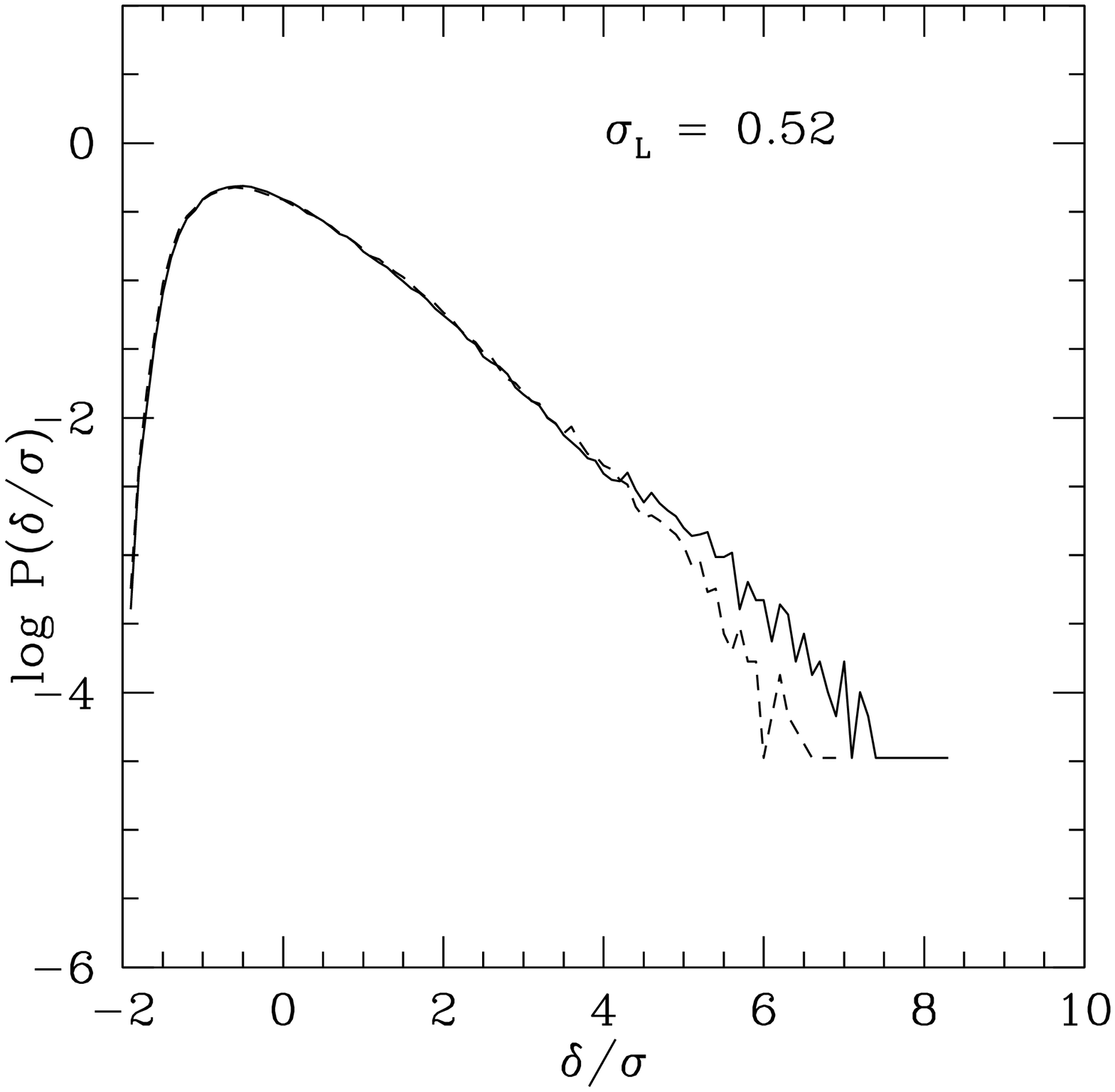}
\epsfxsize=5.7cm \epsfbox{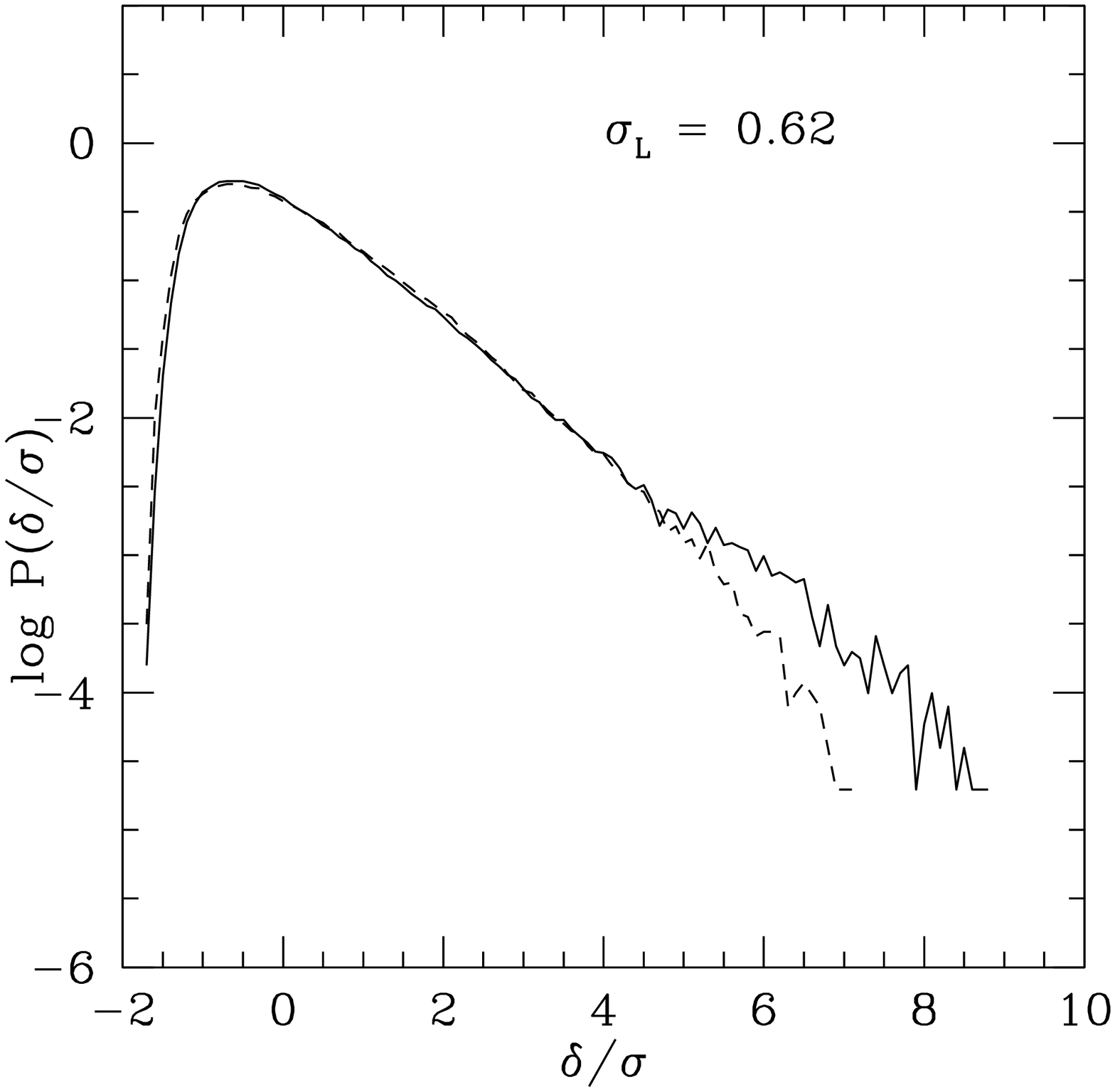}
\epsfxsize=5.7cm \epsfbox{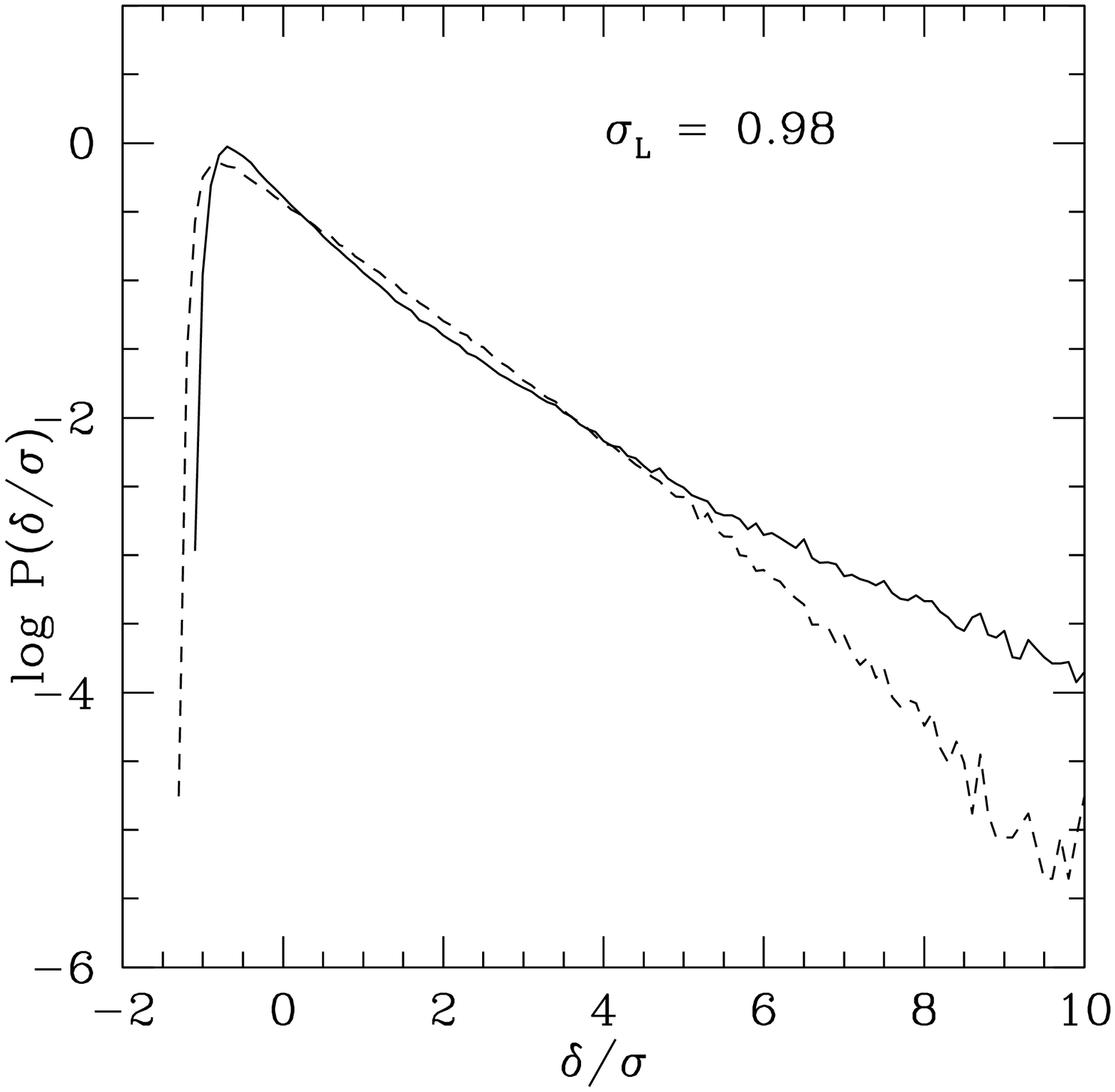}}
\caption[Fig.3]{\label{fig3} A comparison of the real-space PDF,
$P[\delta/\sigma]$ (solid curve) with the redshift-space
PDF, $P\left[\delta_{(z)}/\sigma_{(z)}\right]$ (dashed curve) in SCDM
simulations
at the indicated values of $\sigma_L$. Error bars have been suppressed for
clarity.}
\end{figure*}

\begin{figure*}
\centering\centerline{
\epsfxsize=5.7cm \epsfbox{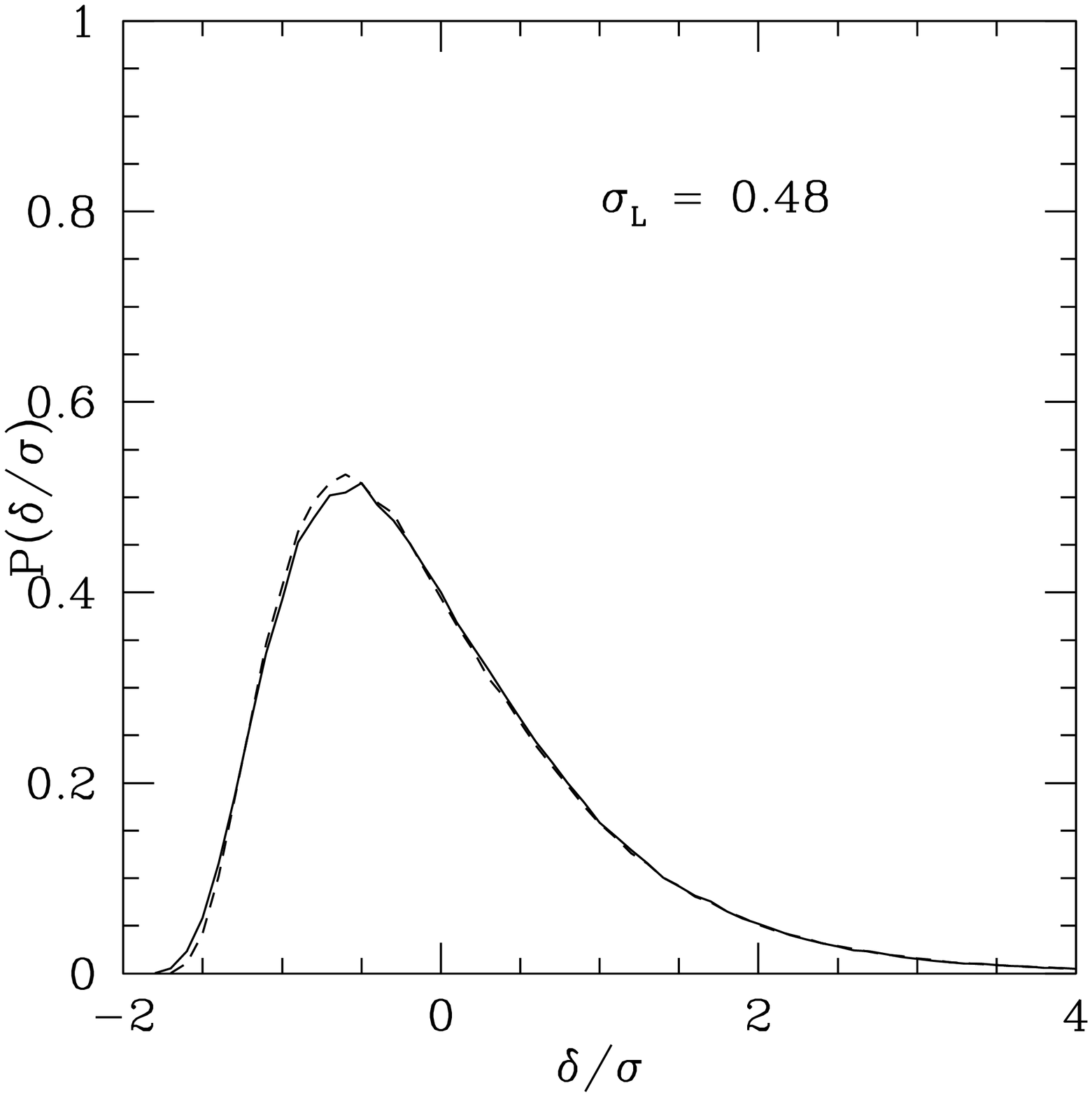}
\epsfxsize=5.7cm \epsfbox{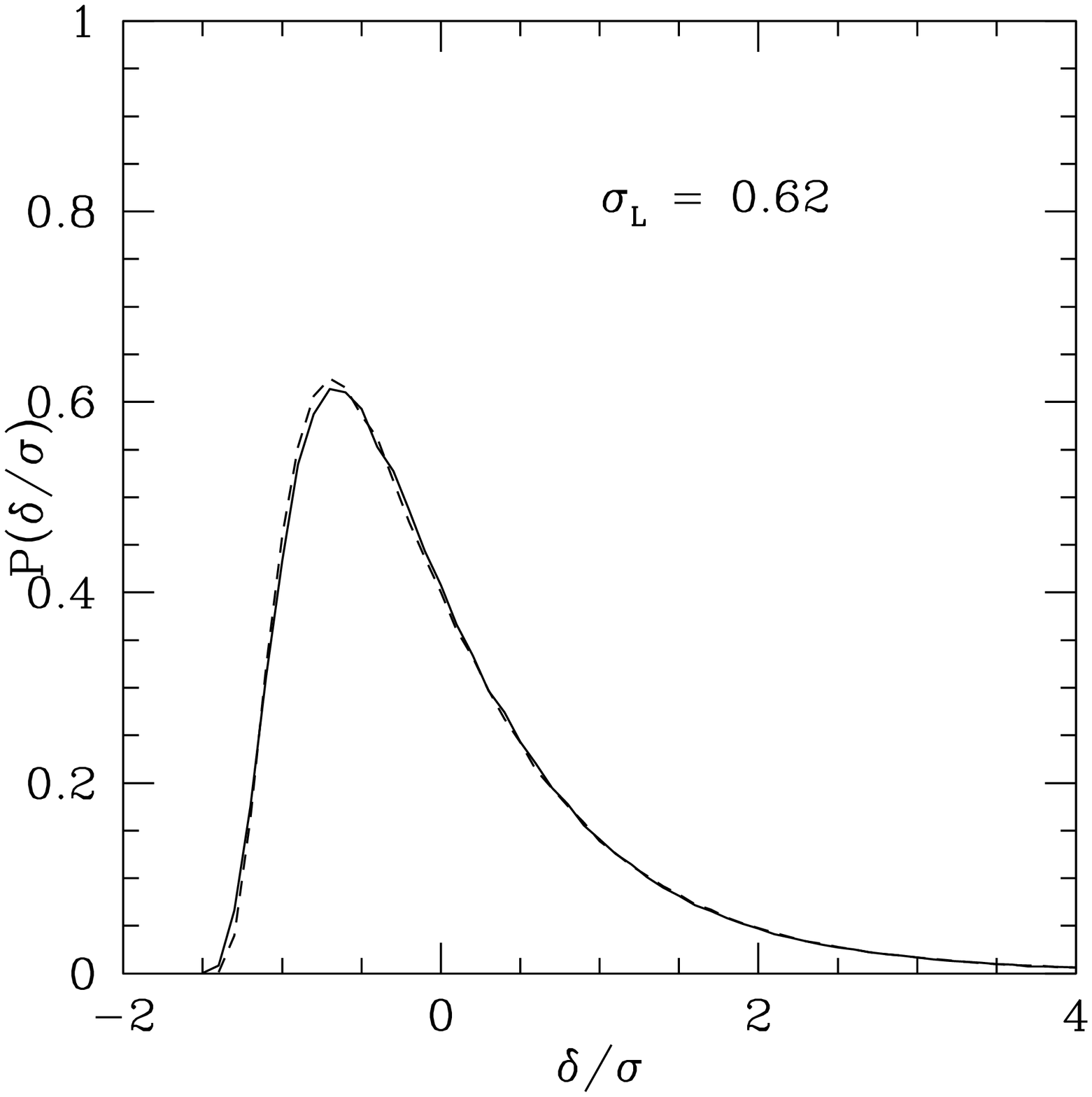}
\epsfxsize=5.7cm \epsfbox{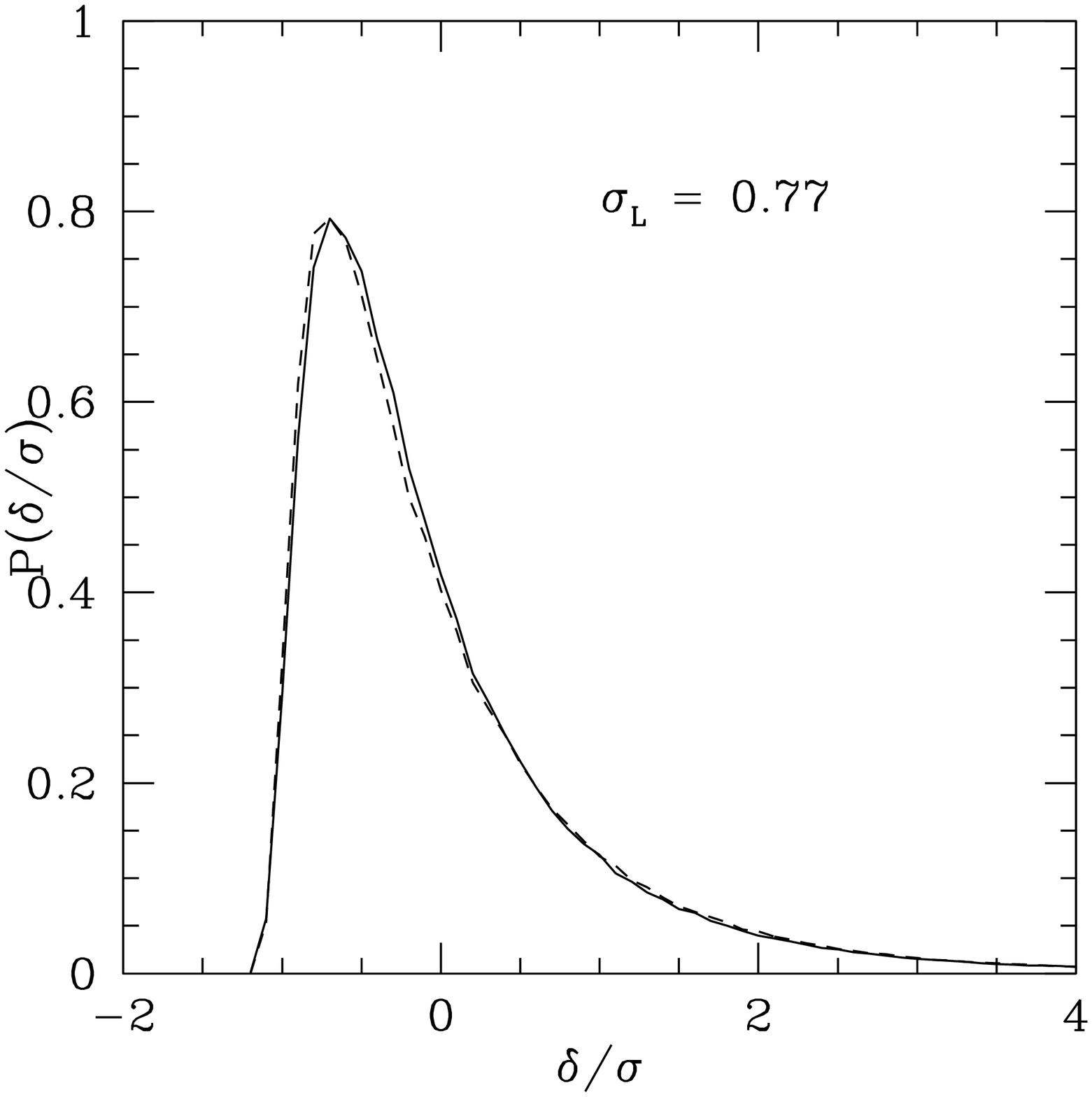}}
\centerline{\epsfxsize=5.7cm \epsfbox{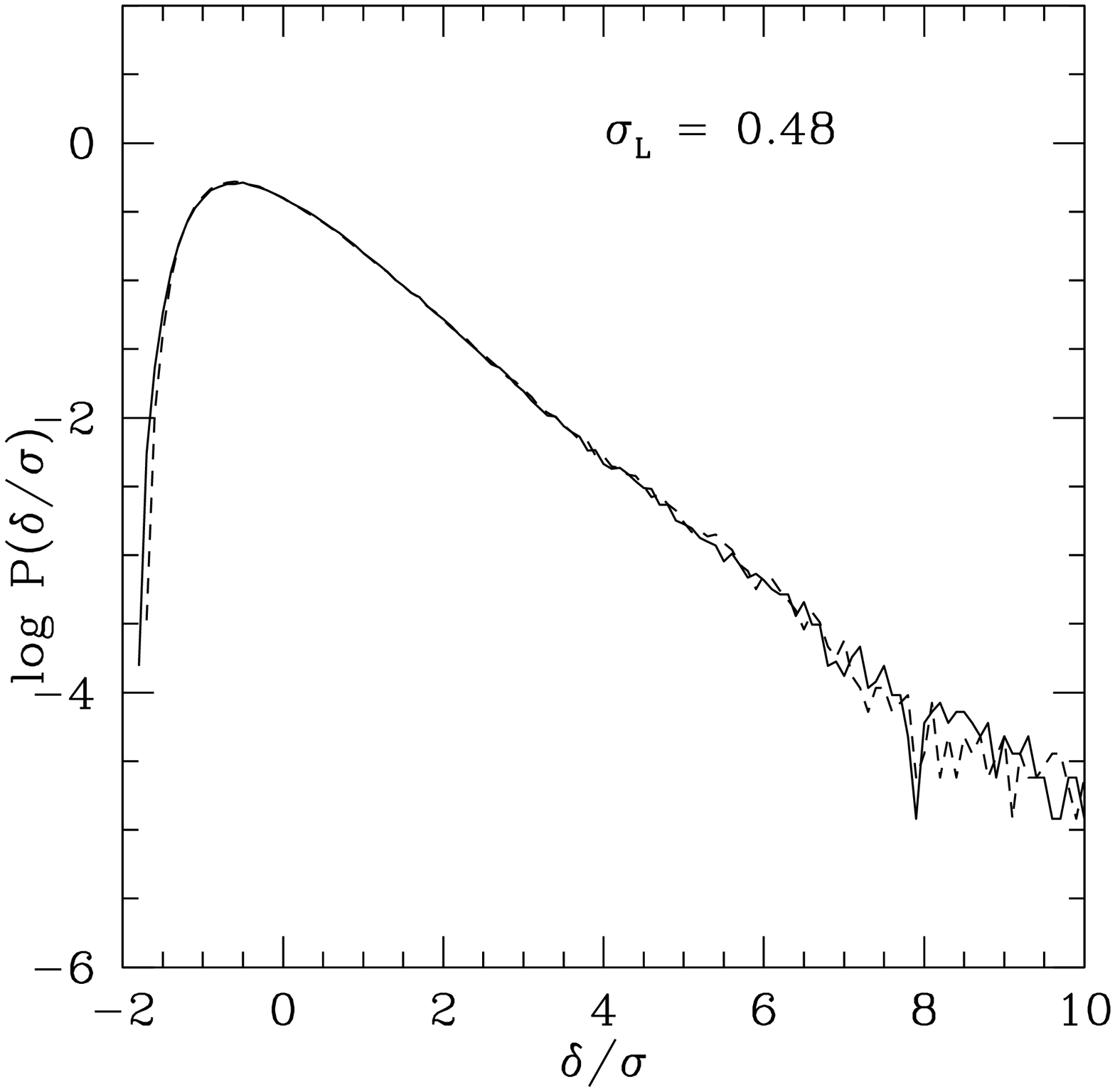}
\epsfxsize=5.7cm \epsfbox{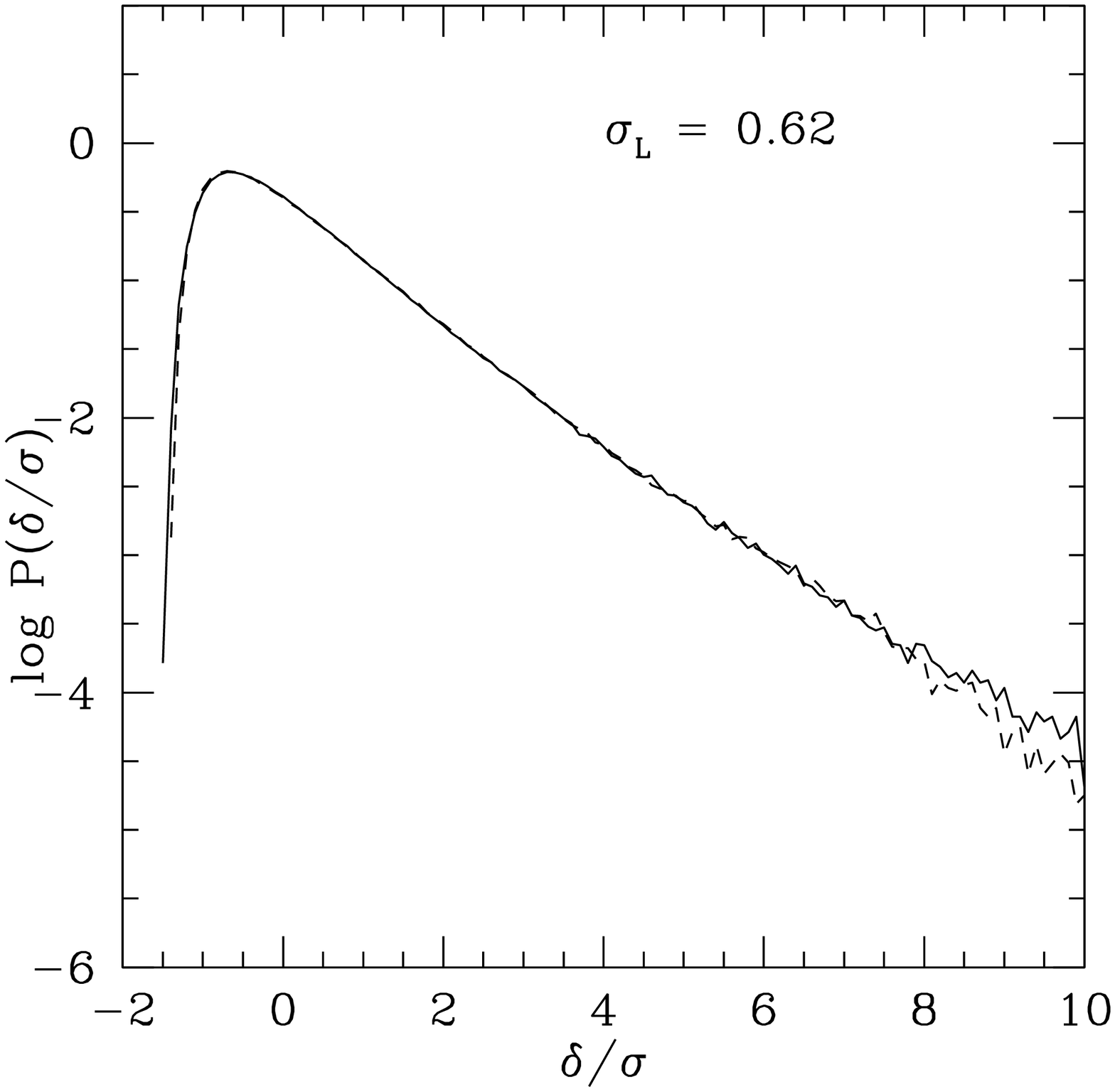}
\epsfxsize=5.7cm \epsfbox{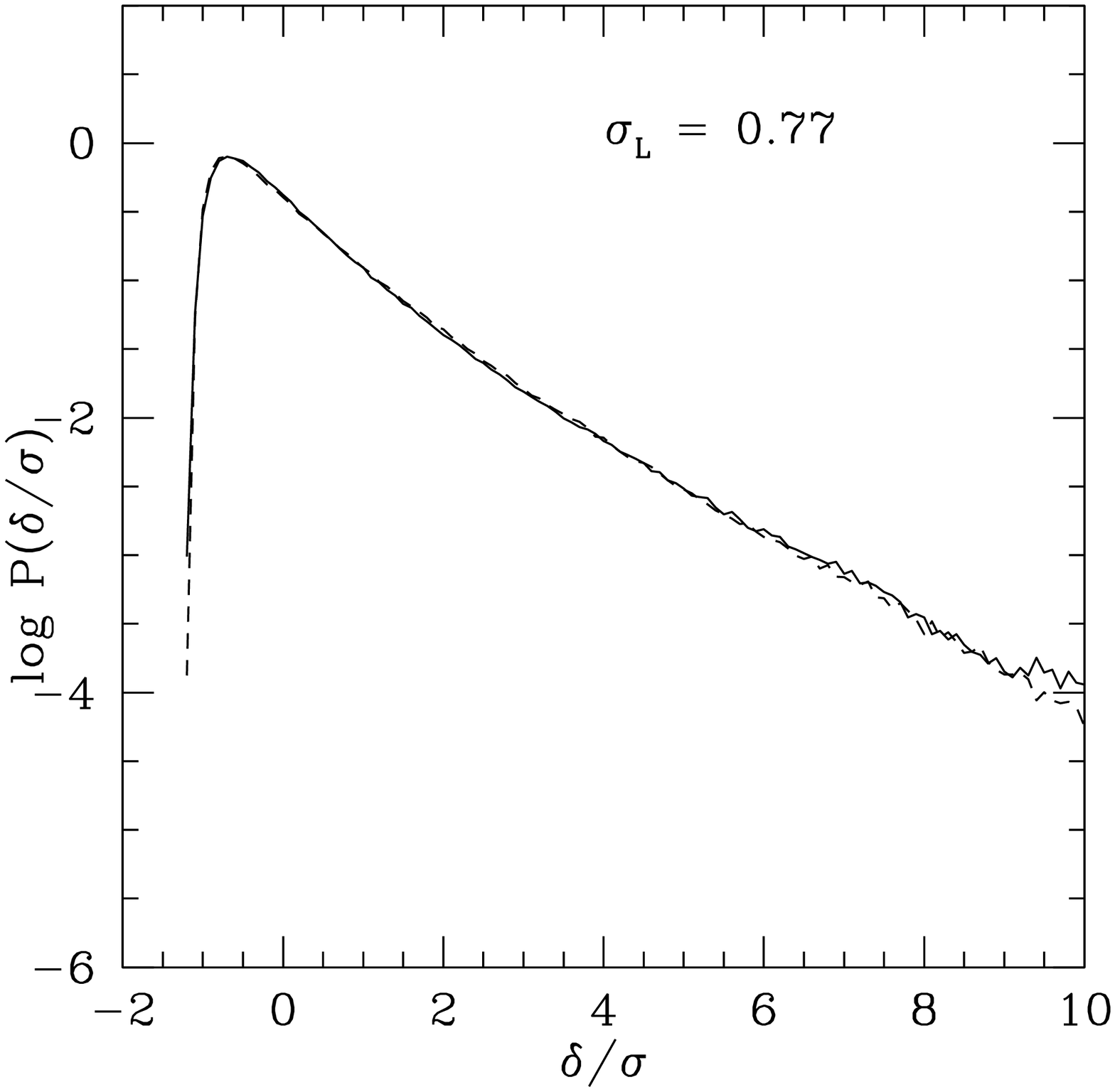}}
\caption[Fig.4]{\label{fig4} A comparison of the real-space PDF,
$P[\delta/\sigma]$ (solid curve) with the redshift-space
PDF, $P\left[\delta_{(z)}/\sigma_{(z)}\right]$ (dashed curve) in flat $\Lambda$CDM
simulations
at the indicated values of $\sigma_L$. Error bars have been suppressed for clarity.}
\end{figure*}

For $\sigma_L \sim 1$,
the SC approximation does a much better job of predicting the
real-space PDF than the redshift-space PDF.
Hence,
we are motivated to consider other possible ways of predicting the latter.
Note that when we normalize the distributions to $\mu \equiv \delta/\sigma$,
the real and redshift space moments are very similar to leading order.
The variances are identical by construction and we have seen that
the dimensionless skewness $B_3$ and kurtosis $B_4$ are quite similar
at leading order. Thus one would expect that when $\sigma_L \rightarrow 0$
the dimensionless redshift PDF would be similar to that in redshift space.
Note that $\mu \equiv \delta/\sigma$ and the dimensionless moments
are the natural variables  to quantify the amount of non-Gaussian effects 
on the PDF, as it is evident from the Edgeworth and Gamma expansions
(e.g., see Gazta\~naga, Fosalba \& Elizalde 2000).

In Fig. \ref{fig3}, we provide a comparison of $P[\delta/\sigma] d[\delta/\sigma]$
in real and redshift space.  It is obvious from these figures that
the agreement is excellent for $\sigma_L \simlt 0.6$ and
only breaks down at $\sigma_L \approx 1$.  
For $\sigma_L \simlt 0.6$, the major difference
occurs at the large-$\delta$ tails of the distributions,
indicating that the higher order 
terms are becoming important.  In particular,
the real and redshift-space distributions diverge at $\delta/\sigma > 4$.

Note nevertheless that the agreement seems better than one would
expect. 
Consider, for example, the values of the variance, skewness and kurtosis
at $R=15$ Mpc/$h$ given in Table 1.
\begin{table}
\caption{The real-space and redshift-space
values of the variance, skewness, and kurtosis measured in
SCDM simulations for a smoothing scale of $R=15$ Mpc/$h$.}
\begin{center}
\begin{tabular}{c|c|c|c|c|c}
& $\sigma$ & $S_3$ & $S_4$ & $B_3$ & $B_4$
\\ \hline \hline 
real &  0.482 & 2.66 & 11.6 & 1.28 & 2.69\\
$z$   & 0.510 & 2.30 &  6.8 & 1.13 & 1.77  \\ \hline 
\end{tabular}
\end{center}
\end{table}
We see that $\sigma_{(z)}^2/\sigma^2$ lies below the 
linear theory prediction in Eq. (\ref{kaiser}) (or
the SC prediction in Eq. \ref{sigmaz}),
while $S_{3(z)}$ and $S_{4(z)}$ are
smaller than the corresponding real space values: in the opposite
direction to the leading order prediction in Eq. (\ref{s3zs4z}).
These two effects seem to cancel each other so that
the values of the dimensionless skewness $B_3$ and kurtosis $B_4$ 
in real and redshift space are quite similar. This suggests
some additional cancellation of the redshift distortions which is
not accounted for by the SC model, most likely due to
nonlinear effects.

Because of the potential significance of this result, we have
repeated this calculation for a flat $\Lambda$CDM model with
$\Omega_\Lambda = 0.8$ (while this value for $\Omega_\Lambda$
is somewhat larger than that suggested by recent observations,
this model and the SCDM model should serve to bracket the likely
effect of redshift distortions).  Our results are displayed in Fig. \ref{fig4}.
Again, we see that, to a good approximation, $P[\delta/\sigma]
d(\delta/\sigma)$ is identical in real and redshift space.
The agreement here is even better than for the SCDM model;
the two distributions are nearly identical for $\sigma$ as large as
0.8.

\section{Conclusions}

The spherical collapse (SC) approximation for the PDF
can provide good agreement
with both the real-space and redshift-space PDF from
numerical SCDM simulations, but it fails at a much smaller
value of $\sigma_L$ in the latter case.  In real space, we
get good agreement with the numerical results for $\sigma_L \simgt 1$,
while in redshift space, good agreement can be obtained
only for $\sigma_L \simlt 0.4$.  This is most likely due
to nonlinear effects in the redshift-space PDF. An interesting
advantage of the SC approach to derive the PDF is the possibility
of extending the results  to the case of
non-Gaussian initial conditions (see also Gazta\~naga \& Fosalba 1998,
Gazta\~naga \& Croft 1999). We only need to replace $P_0$ in 
equation (\ref{PDF1}) by the corresponding non-Gaussian initial PDF.

We also find an unexpected result in our simulations:
the redshift space PDF, $P[\delta_{(z)}]$, is, to a good approximation,
a simple rescaling of the real space PDF, $P[\delta]$, i.e.,
$P[\delta/\sigma] d[\delta/\sigma] = P\left[\delta_{(z)}/\sigma_{(z)}\right]
d\left[\delta_{(z)}/\sigma_{(z)}\right]$.  This rescaling provides an excellent
approximation to the redshift-space PDF for $\sigma_L < 1$
in both the SCDM and $\Lambda$CDM models.
We have not found an entirely satisfactory explanation for this.
While it would
be an obvious result from linear theory, its validity extends well
beyond the linear regime.
For example, this rescaling applies very well to the regime for which $P[\delta]$ is highly non-Gaussian and
the Kaiser prediction for
$\sigma_{(z)}$ from linear theory (Eq. \ref{kaiser}) breaks down.
This result seems to stem from a cancellation between the linear
effect given by Eq. (\ref{redshift}) and the nonlinear redshift
distortions.  It is perhaps ironic that this simple-minded
rescaling gives a much better prediction for the redshift-space
PDF than our sophisticated perturbation-theory approximation.

\section*{Acknowledgments}

We thank L. Hui, J. Frieman, and R. Scoccimarro for helpful discussions. E.G.
acknowledge support by grants from IEEC/CSIC and DGES(MEC)(Spain) project
PB96-0925, and Acci\'on Especial ESP1998-1803-E.  R.J.S. was supported in part
by the Department of Energy (DE-FG02-91ER40690).  R.J.S.  is grateful for the
hospitality of the Fermilab Theoretical Astrophysics Group, where this work
was initiated.

\end{document}